\documentclass[%
 reprint,
 table,
superscriptaddress,
nofootinbib,
 amsmath,amssymb,
 aps,
 pra,
]{revtex4-2}

\usepackage{graphicx}% Include figure files
\usepackage{dcolumn}% Align table columns on decimal point
\usepackage{bm}% bold math
\usepackage[caption=false, position=top]{subfig}
\usepackage[capitalise]{cleveref}
\usepackage{underscore}
\usepackage{comment}

\usepackage{textgreek}
\usepackage{amsmath}
\usepackage{mathtools}
\usepackage{braket}
\usepackage{qcircuit}

\DeclareMathOperator{\Tr}{Tr}

\usepackage{amssymb}
\usepackage{pifont}
\usepackage[dvipsnames]{xcolor}
\usepackage{adjustbox}

\newcommand{\xmark}{\ding{55}}%
\newcommand{\greencheck}{{\color{LimeGreen}\checkmark}}
\newcommand{\redx}{{\color{Red}\xmark}}

\begin{document}

\preprint{APS/123-QED}

\title{Characterization of entanglement on superconducting quantum computers of up to 414 qubits}% Force line breaks with \\

\author{John F Kam}
\thanks{john.kam@monash.edu}
\author{Haiyue Kang}
\affiliation{%
School of Physics, University of Melbourne, VIC, Parkville, 3010, Australia.
}%
\author{Charles D Hill}
\affiliation{%
School of Physics, University of Melbourne, VIC, Parkville, 3010, Australia.
}%
\affiliation{%
School of Mathematics and Statistics, The University of Melbourne, Parkville, 3010, Australia.
}%
\author{Gary J Mooney}
\author{Lloyd C L Hollenberg}
\affiliation{%
School of Physics, University of Melbourne, VIC, Parkville, 3010, Australia.
}%

\date{\today}% It is always \today, today,

\begin{abstract}
As quantum technology advances and the size of quantum computers grow, it becomes increasingly important to understand the extent of quality in the devices. As large-scale entanglement is a quantum resource crucial for achieving quantum advantage, the challenge in its generation makes it a valuable benchmark for measuring the performance of universal quantum devices. In this work, we study entanglement in Greenberger-Horne-Zeilinger (GHZ) and graph states prepared on the range of IBM Quantum devices. We generate GHZ states and investigate their coherence times with respect to state size and dynamical decoupling techniques. A GHZ fidelity of $0.519 \pm 0.014$ is measured on a 32-qubit GHZ state, certifying its genuine multipartite entanglement (GME). We show a substantial improvement in GHZ decoherence rates for a 7-qubit GHZ state after implementing dynamical decoupling, and observe a linear trend in the decoherence rate of~$\alpha=(7.13N+5.54)10^{-3}$ \textmu s$^{-1}$ for up to~$N=15$ qubits, confirming the absence of superdecoherence. Additionally, we prepare and characterize fully bipartite entangled native graph states on 22 superconducting quantum devices with qubit counts as high as 414 qubits, all active qubits of the 433-qubit IBM Osprey device. Analysis of the decay of 2-qubit entanglement within the prepared states shows suppression of coherent noise signals with the implementation of dynamical decoupling techniques. Additionally, we observe that the entanglement in some qubit pairs oscillates over time, which is likely caused by residual ZZ-interactions. Characterizing entanglement in native graph states, along with detecting entanglement oscillations, can be an effective approach to low-level device benchmarking that encapsulates 2-qubit error rates along with additional sources of noise, with possible applications to quantum circuit compilation. We develop several tools to automate the preparation and entanglement characterization of GHZ and graph states. In particular, a method to characterize graph state bipartite entanglement using just 36 circuits, constant with respect to the number of qubits.

\end{abstract}

\maketitle

\section{Introduction}

\begin{table*}[t]
    \normalsize
    \centering
    \renewcommand{\arraystretch}{1.2}
    \begin{tabular}{p{.11\textwidth} p{.6\textwidth} p{.14\textwidth} p{.11\textwidth}}
        \hline\hline
        Entangled State &  Experiment & Device(s) & Section\\ \hline
        GHZ \newline (static) & GHZ state generation and verification up to 32 qubits & \textit{ibm\texttt{\_}\!washington} \newline (127 qubits)& Section \ref{section:ghzsize} \\
        GHZ \newline (dynamic) & Extending 7-qubit GHZ state lifetimes with dynamical decoupling & \textit{ibmq\texttt{\_}\!mumbai} \newline (27 qubits) & Section \ref{subsection:ghzpdd} \\
        GHZ \newline (dynamic) & Scaling of GHZ decoherence rate versus state size up to 15 qubits & \textit{ibm\texttt{\_}\!hanoi} \newline (27 qubits) & Section \ref{subsection:ratescaling} \\ \hline
        Graph state \newline (static) & Scalable whole-device bipartite entanglement characterization up to 414 qubits & 22 devices, \newline see \cref{tab:summary} & Section \ref{section:graphstate} \\ 
        Graph state \newline (dynamic) & Testing dynamical decoupling for 127-qubit whole-device bipartite entanglement & \textit{ibm\texttt{\_}\!brisbane} \newline (127 qubits) & Section \ref{section:graphstatedecay} \\
        \hline\hline
        
    \end{tabular}
    \caption{Guiding table summarizing all main experiments, the devices tested, and the relevant sections of the paper.}
    \label{tab:guiding}
\end{table*}

Producing large-scale entangled states on many qubits represents an important frontier in the race to scale up physical quantum computers. In quantum computers, entanglement manifests as non-classical correlations between qubits, such that qubits involved in an entangled system cannot be described independently from each other \cite{Horodecki2009QuantumEntanglement,Einstein1935CanComplete}. Entanglement has been found to play a significant role in achieving quantum advantage \cite{Jozsa2003OnSpeed-up, Vidal2003EfficientComputations, Datta2005EntanglementQubit, Kendon2006EntanglementAlgorithm}, and many quantum information processing protocols explicitly rely on entanglement as a resource \cite{Ekert1991QuantumTheorem,Bennett1992CommunicationStates,Bennett1993TeleportingChannels,Wootters1998QuantumResource}. Furthermore, entanglement over large arrays of qubits is essential in many fault-tolerant computation schemes \cite{Raussendorf2006AComputer,Brun2006CorrectingEntanglement,Fowler2012SurfaceComputation}. Multipartite entanglement is known for its complex structure \cite{Horodecki2009QuantumEntanglement}, and modern quantum devices have passed the number of qubits a classical computer can store an arbitrary quantum state on \cite{Chow2021IBMBarrier,Madsen2022QuantumProcessor}. Consequently, as noisy intermediate-scale quantum (NISQ) devices \cite{Preskill2018QuantumBeyond} continue to improve in both size and error rates, it becomes paramount to characterize the ability of a quantum computer to generate and maintain large entangled states in a scalable manner.

Verifying multipartite entanglement on a quantum device requires measuring the fidelity of a highly entangled multi-qubit state. Greenberger-Horne-Zeilinger (GHZ) states \cite{Greenberger1989GoingTheorem} are a popular choice, as measuring a GHZ state fidelity of greater than 0.5 is sufficient for verifying genuine multipartite entanglement (GME). For quantum devices with full qubit control, GHZ states of sizes 27 and 29 qubits have been observed on superconducting systems \cite{Mooney2021GenerationComputer,Yang2022EfficientDevices}, and GHZ states of size 32 qubits have been observed on ion-trap systems \cite{Moses2023AProcessor}. Graph states, also known as cluster states, are another widely studied class of multipartite entangled states. Graph states are useful for showing mixed state bipartite entanglement, and full bipartite entanglement has been observed on up to 65-qubit graph states in superconducting systems \cite{Wang201816-qubitEntangled,Mooney2019EntanglementComputer,Mooney2021Whole-deviceComputer}, and 20-qubit graph states in ion-trap systems \cite{Friis2018ObservationSystem}. There has also been recent work showing violation of robust Bell inequalities for 57-qubit path graph states \cite{Yang2022TestingDevices}, and genuine entanglement for 51-qubit path graph states and 30-qubit 2D graph states \cite{Cao2023GenerationQubits} on superconducting devices.

Increasing the size of multipartite entanglement is profitless if the entangled state degrades too rapidly to be able to perform meaningful operations on. A source of concern when engineering physical quantum devices is \textit{superdecoherence}; a phenomenon where qubit decoherence rates scale linearly with the number of qubits due to the coupling of multiple qubits with a single reservoir \cite{Kattemolle2020ConditionsSuperdecoherence}.  Studying the decay of entangled states can reveal information about the noise underlying a quantum system. For example, the decoherence of GHZ states has been used to show superdecoherence (or the lack thereof) in superconducting and ion-trap systems \cite{Monz201114-qubitCoherence,Ozaeta2019DecoherenceProcessor}. 

Techniques such as dynamical decoupling have been developed to suppress the effect of environmental noise on quantum states. Dynamical decoupling, a quantum control technique that employs sequences of control pulses to suppress the effect of environmental noise on quantum states, has been shown to be remarkably effective at protecting four-qubit GHZ states on ion-trap devices \cite{Kaufmann2017ScalableEntanglement}.

In this work, we investigate both GME in the form of GHZ state fidelities and mixed bipartite whole-device entanglement using native-graph states. We develop several tools to automate the preparation and verification of these states over the range of IBM Quantum devices. In particular, we employ an automated GHZ state embedding scheme that embeds tree-type GHZ preparation circuits with minimum-depth on heavy-hexagonal qubit architectures to generate up to 32-qubit GHZ states on the 127-qubit \textit{ibm\texttt{\_}\!washington} device. Using multiple quantum coherences (MQC) \cite{Wei2020VerifyingCoherences}, we record a fidelity of $0.519\pm 0.014$ for the 32-qubit state after mitigating for measurement errors. 

We then test dynamical decoupling-based schemes in preserving GHZ state coherences for a 7-qubit state, and found both periodic dynamical decoupling (PDD) and a \textpi -pulse to be highly effective. From then on incorporating a \textpi -pulse, we investigate how GHZ state decoherence rates scale with number of qubits, and observe a linear trend up to 15-qubit GHZ states. For graph states, we develop a bipartite entanglement characterization protocol focusing on adaptability and scalability. The protocol first prepares a native-graph state on the device and then performs quantum state tomography on each pair of qubits in parallel (projected onto maximally entangled Bell pairs by measurements on their neighbours) in order to calculate pair-wise negativities. Using the procedure, we find whole-device entanglement---i.e. the entangled pairs of qubits form a connected entanglement graph that includes every device qubit--on 21 IBM Quantum devices, including three 127-qubit systems. We further show entanglement across 414 qubits on the 433-qubit Osprey device. 

We finally investigate dynamical decoupling for preserving native-graph state entanglement. Notably, we observe revivals in entanglement signals for several qubit pairs. The observed resurgent signals in negativity are consistent with signals produced by residual ZZ interactions, which are known to affect superconducting transmon qubits and generate local entanglement  \cite{McKay2019Three-QubitBenchmarking,Magesan2020EffectiveGate,Ku2020SuppressionSystem,Sundaresan2020ReducingEchoes,Zhao2021SuppressionProcessor}. We observe improvement in mean device entanglement lifetimes after implementing PDD.

To enhance the paper's navigability, we provide a guiding table in \cref{tab:guiding} summarizing all main experiments, the devices tested, and the relevant sections of the paper.

\section{Generation and Decay of Greenberger-Horne-Zeilinger States}\label{section:ghzsize}
\subsection{GHZ states and verifying genuine multipartite entanglement}

\begin{figure}
    \centering
    \includegraphics[width=0.85\columnwidth]{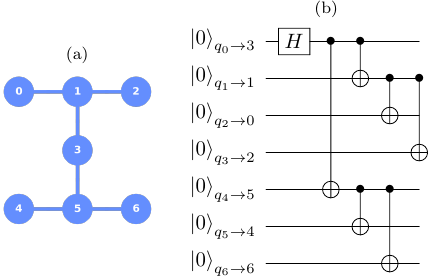}
    \caption{GHZ state preparation circuit on a 7-qubit processor. \textbf{(a)} Physical layout of the 7-qubit Falcon R5.11H processor, where nodes represent qubits and edges represent potential two-qubit operations. \textbf{(b)} Optimal GHZ circuit embedding on (a), where qubits are mapped in a way that avoids the need for SWAP operations. The circuit maximizes the number of CNOTs performed in parallel, minimizing the overall CNOT depth to four.\label{fig:ghz1}}
\end{figure}

GHZ states \cite{Greenberger1989GoingTheorem} are highly entangled multipartite states that are uniquely fragile to noise, where a single-qubit phase error can destroy the whole-state entanglement. The ability of a quantum device to generate large GHZ states with high fidelity depends on a holistic combination of factors, including qubit count, gate error rates, coherence times, and cross-talk. Generating and verifying such states has therefore become a valuable benchmark for NISQ devices. GHZ states are typically prepared by initializing a source qubit in the $\ket{+}$ state, and then iteratively applying CNOT gates from the source qubit (or any other qubit that has already had a CNOT applied in this manner) to all other qubits prepared in the $\ket{0}$ state (\cref{fig:ghz1}). The resulting state is an equal superposition of all subsystems in the $\ket{0}$ state and all subsystems in the $\ket{1}$ state. Formally, an $N$-qubit GHZ state is expressed as:
\begin{equation}
    \ket{\mathrm{GHZ}_N} = \frac{\ket{0}^{\otimes N} + \ket{1}^{\otimes N}}{\sqrt{2}}
\end{equation}

To certify \(N\)-qubit GME, it is sufficient to show a GHZ fidelity of greater than 0.5 \cite{Leibfried2005CreationState} (note that a GHZ state may be GME with a fidelity under 0.5). The fidelity \(F\) between a pure target state \(\rho^{\mathrm{ideal}}\) and the actual (noisy) state \(\rho\) is calculated as
\begin{equation}\label{eq:fidelity1}
    F(\rho, \rho^{\mathrm{ideal}}) = \Tr({\rho\rho^{\mathrm{ideal}}})
\end{equation}
The resource requirement for full quantum state tomography scales exponentially with the number of qubits, making it intractable to obtain $\rho$ using this approach even for medium-sized systems. Fortunately, GHZ states of all sizes have the convenient property that their density matrices (ideally) consist of only four non-zero corner elements. As a result, for GHZ states, \cref{eq:fidelity1} can be re-expressed as
\begin{equation}
    F_{\mathrm{GHZ}} = \Tr({\rho\rho_{\mathrm{GHZ}}})= \frac{P + C}{2}
\end{equation}
where \(P = \rho_{00\dots 0,00\dots 0} + \rho_{11\dots1,11\dots 1}\) are the directly measured populations, and \(C = |\rho_{00\dots 0,11\dots 1}| + |\rho_{11\dots 1,00\dots 0}|\) are the coherences which can be measured using either parity oscillations \cite{Leibfried2005CreationState, Monz201114-qubitCoherence} or multiple quantum coherences (MQC) \cite{Wei2020VerifyingCoherences}. 

MQC is advantageous due to its natural integration with dynamical decoupling-based techniques, such as the Hahn echo, which refocuses noise and mitigates dephasing, as well as readout error mitigation, since readout errors are typically the dominant noise factor for low-depth circuits. MQC has been used to verify GHZ states of up to 29 qubits on superconducting quantum devices \cite{Wei2020VerifyingCoherences, Mooney2021GenerationComputer, Yang2022EfficientDevices}. The methodology for computing GHZ coherences via MQC can be summarized as follows:
\begin{enumerate}
    \item Prepare the $N$-qubit GHZ state in the form $\frac{1}{\sqrt{2}}(\ket{00\dots 0} + \ket{11\dots 1})$ as exemplified in \cref{fig:ghz1}(b).
    \item (Optionally) Apply a refocusing $\pi$-pulse, i.e., an $X$-gate on each qubit.
    \item Apply a phase rotation of $\phi$ on each qubit in the GHZ state, adding an accumulative phase of $N\phi$ to the overall state: $\frac{1}{\sqrt{2}}(\ket{00\dots 0} + e^{-iN\phi}\ket{11\dots 1})$.
    \item Disentangle the state by applying the inverse of the GHZ state preparation circuit from step 1. The accumulated phase is mapped onto qubit 0: $\frac{1}{\sqrt{2}}(\ket{0} + e^{-iN\phi}\ket{1}) \otimes \ket{00\dots 0}$.
    \item Obtain the overlap signal $S_\phi$ as the probabilities of measuring the $\ket{00\dots 0}$ state for different values of $\phi$.
\end{enumerate}
The $N$-qubit GHZ coherence can then be calculated as $C = 2\sqrt{I_N}$ where $I_N$ are the overlap signal amplitudes which can be obtained by applying a Fourier transform to $S_\phi$:
\begin{equation}\label{eq:signal amplitude}
    I_q = n^{-1}\left\lvert\sum_\phi e^{iq\phi}S_\phi\right\rvert
\end{equation}
where $n$ is the number of angles $\phi$ in the summation. To detect up to frequency $N+1$, we measure $S_\phi$ for angles \(\phi = \frac{\pi j}{N+1}\), \(j = 0, 1, \dots, 2N+1\).
Further details including deriving the fidelity from the overlap signal concretely is presented in \cite{Wei2020VerifyingCoherences,Mooney2021GenerationComputer}.

\subsection{GHZ state embedding on physical topologies}
When preparing GHZ states on physical devices, accounting for hardware topology and gate error rates is crucial for maximizing the final state fidelity. In previous experiments \cite{Mooney2021GenerationComputer}, circuit embedding was performed manually. While suitable for smaller devices, the introduction of 433-qubit processors, and most recently, the announcement of a 1121 qubit processor \cite{Castelvecchi2023IBMChip} necessitate methods for automated embedding of device entangling circuits.

In this work, we develop a topology-agnostic GHZ state preparation scheme that constructs tree-type GHZ preparation circuits on heavy-hexagonal layouts with minimal depth. The method additionally involves selecting qubit subsets with optimized parameters such as low two-qubit gate error rates. The exact parameter or combination of parameters to optimize is specified by assigning corresponding weights to the edges of the graph. We initially considered only the CNOT error rate. Our algorithm is divided into two components. The first component embeds a GHZ circuit as a directed tree branching out from a single source qubit. We apply CNOT gates in parallel prioritizing least depth first and lowest two-qubit error rate second. The second component runs the first component algorithm multiple times, trialling each physical qubit or a subset of physical qubits as the source qubit. We then select the circuit with the least depth as the primary criterion and the lowest total cost parameter as the secondary criterion. This way, we can conveniently embed least-depth GHZ states on larger and larger quantum devices.
\begin{figure}[t]
    \centering
    \includegraphics[width=\columnwidth]{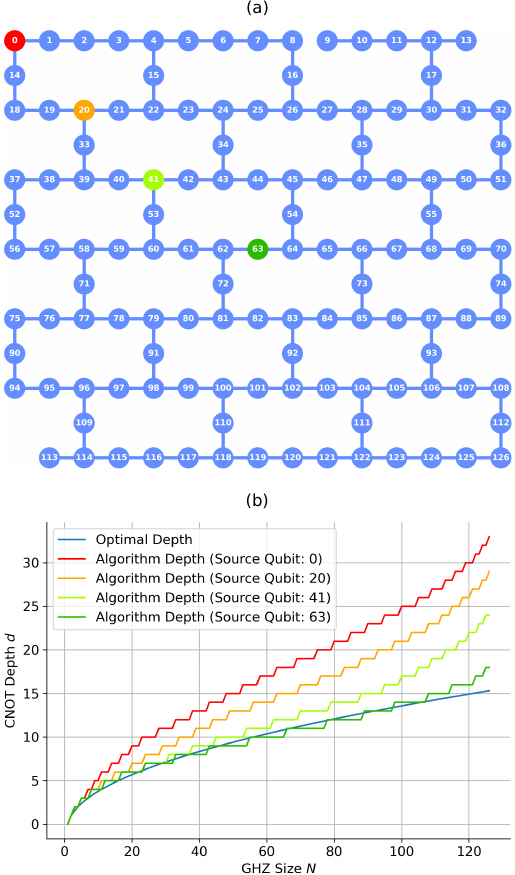}
    \caption{The physical qubit layout of a 127-qubit Eagle processor \textbf{(a)} and the corresponding plot for algorithmically embedded GHZ circuit depth vs. GHZ state size \textbf{(b)}. The optimal depth curve plots \cref{eq:heavy hex size} which describes the maximum number of qubits in a GHZ state $N$ that can be prepared with a circuit of depth $d$. The remaining curves showcase the performance of the GHZ embedding algorithm for different source qubits (color coded) as highlighted in (a). \label{fig:ghzdepthscaling}}
\end{figure}
\begin{figure*}
    \centering
    \includegraphics[width=0.75\textwidth]{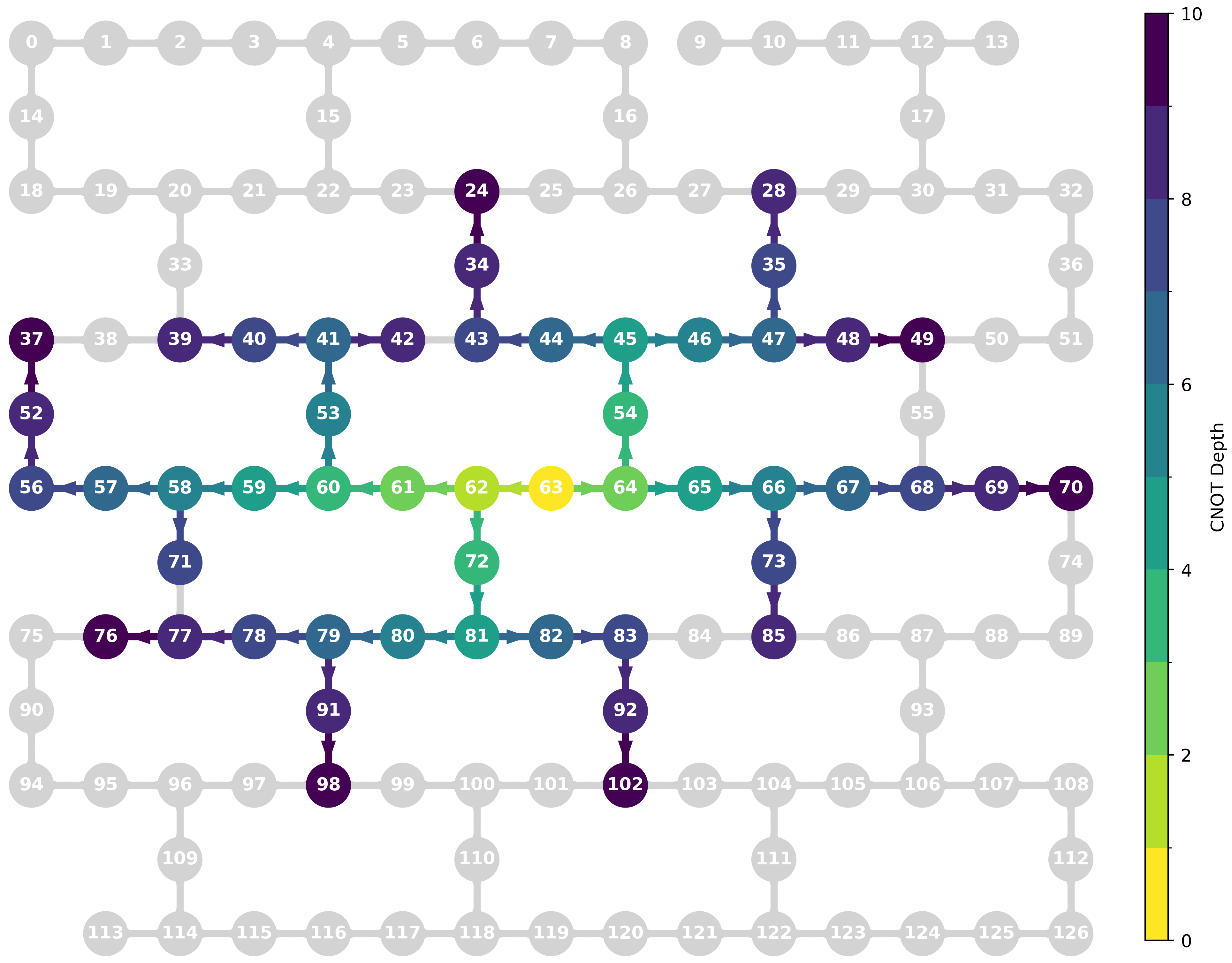}
    \caption{GHZ state of size 60 qubits algorithmically embedded on \textit{ibm\texttt{\_}\!washington}. Qubit 63 serves as the initial source qubit, with arrows indicating the direction of CNOT gates (control $\to$ target qubit) and the color bar mapping the CNOT depths.\label{fig:ghzgen washington map}}
\end{figure*}

IBM employs a heavy-hexagonal lattice as its principal architecture for all their devices. The choice in topology is motivated by a reduction in qubit frequency collisions \cite{Hertzberg2021Laser-annealingProcessors} and spectator errors \cite{Takita2017ExperimentalQubits}, as well as surface code versatility \cite{Chamberland2020TopologicalQubits}. It has been shown for star graph states (which are equivalent to GHZ states under LOCC) that the CNOT depth to construct states of size $N$ scales as approximately \(\sqrt{2N}\) on heavy-hex architectures \cite{Yang2022TestingDevices}. Precisely, a GHZ circuit embedded on an infinite heavy-hexagonal lattice with depth $d$ can prepare states of up to size
\begin{equation}\label{eq:heavy hex size}
    N = \frac{d(d+1)}{2} + 1
\end{equation}
where $N$ is the number of qubits.

For the heavy-hex topology, our algorithm embeds GHZ states with this optimal depth scaling up to a limit imposed by the boundaries of the physical device. We illustrate this for a 127-qubit Eagle processor in \cref{fig:ghzdepthscaling}, where it can be seen that a tree-type GHZ state with a centrally located source qubit closely follows \cref{eq:heavy hex size}, whereas selecting source qubits closer to the boundaries results in worse scaling. We showcase an algorithmically embedded example 60-qubit GHZ state on the same processor in \cref{fig:ghzgen washington map}. We note that the algorithm is designed to be compatible with any finite-degree device graph, although its performance on other topologies has not yet been investigated.

\subsection{Scalable quantum readout-error mitigation (M3)}
Readout or measurement errors represent the largest source of noise for low-depth circuits executed on NISQ devices. Readout error rates of even a few percent can be debilitating to the output fidelity of otherwise well-performing systems. Nevertheless, if the readout error takes on a predominantly classical form, which has been shown to be largely true for IBM Quantum transmon devices \cite{Mooney2021GenerationComputer}, its effects on measured probability distributions can be mitigated via post-processing. In its simplest form, quantum readout error mitigation (QREM) solves the linear equation
\begin{equation}\label{eq:qrem}
    \Vec{p}_{\mathrm{noisy}} = \mathbf{A}\Vec{p}_{\mathrm{ideal}}
\end{equation}
where $\Vec{p}_{\mathrm{noisy}}$ is the noisy probability vector returned by the system, $\Vec{p}_{\mathrm{ideal}}$ is the ideal probability vector in the absence of readout errors (but may include other errors), and $\mathbf{A}$ is the \(2^N\times 2^N\) calibration matrix.

There are limitations to QREM, with the most significant one being the exponential scaling of classical resources required to solve for $\Vec{p}_{\mathrm{ideal}}$ with respect to $N$. There are a variety of approaches to overcome this challenge, which often involve approximating the calibration matrix by reducing it to a tensor product of single-qubit components \cite{Mooney2021GenerationComputer,Nation2021ScalableComputers}. More recently, a measurement error mitigation method called M3 (Matrix-Free Measurement Error Mitigation) has demonstrated order-of-magnitude improvement in mitigation time over traditional methods \cite{Nation2021ScalableComputers}. M3 relies on two main optimizations: subspace reduction of the full calibration matrix $\mathbf{A}$ based on noisy input bit strings, and implementation of a matrix-free, iterative method for solving the system of linear equations. For two bit-strings $\mathtt{row}, \mathtt{col} \in \{0,1\}^N$, where qubit 0 corresponds to the least significant bit, the method directly computes matrix elements $\mathbf{A}^{(T)}_{{\mathtt{row}, \mathtt{col}}}$ as
\begin{equation}
    \mathbf{A}^{(T)}_{\mathtt{row}, \mathtt{col}} = \prod_{k=0}^{N-1} P^{(k)}(\mathtt{row}[N-1-k]\to\mathtt{col}[N-1-k]),
\end{equation}
where $P^{(k)}(\mathtt{row}[N-1-k]\to\mathtt{col}[N-1-k])$ corresponds to the probability of the $k$-th qubit being in state $\mathtt{row}[N-1-k]$ and measured in state $\mathtt{col}[N-1-k]$. This error mitigation technique has some caveats. Firstly, M3 works natively with quasi-probability distributions which can contain negative elements. These non-physical probabilities arise from the finite sampling and, while still adding up to one, are incompatible with methods such as MQC. Therefore, a classical algorithm is used to efficiently convert the quasi-probabilities into the closest physical probability distribution under the $L^2$-norm, and it runs in $O(N)$ time \cite{Smolin2012EfficientNoise}. Furthermore, the speedup of the mitigation process depends on the sparsity of the measured probability distributions. A sparser $\Vec{p}_{\mathrm{noisy}}$ corresponds to a greater subspace reduction, which makes M3 optimal for GHZ states since they have only two measurement outcomes. However, scalable mitigation of readout errors comes at the expense of increased uncertainty for measurement outcomes. The mitigation overhead $\mathcal{M}$ is given by:
\begin{equation}
    \mathcal{M} = \lVert \mathbf{A}^{-1}\rVert^2_1
\end{equation}
where \(\lVert X \rVert_1\) is the trace norm of $X$. Quantity $\mathcal{M}$ gives an upper bound to the standard deviation of an observable
\begin{equation}
    \sigma \leq \sqrt{\mathcal{M}/s}
\end{equation}
where $s$ is the number of samples. Thus, results mitigated using M3 will require more samples to achieve similar uncertainty with results mitigated using traditional QREM. 

Nevertheless, the benefits of employing M3 heavily outweigh the limitations in this use case. For our GHZ experiments, we employ M3 using the publicly available \href{https://qiskit.org/documentation/partners/mthree/index.html}{\texttt{mthree}} Python package. We apply the correction algorithm with all the default settings, which include correcting bit strings up to a Hamming distance equal to GHZ size $N$, a convergence tolerance of the iterative method of $10^{-5}$, and a maximum number of iterations of 25.

\subsection{Verifying 32-qubit GHZ states on a 127-qubit Eagle Processor}

We prepare and measure the fidelities of tree-type GHZ states of sizes $N = 27, 28, 29, 30, 31, 32$, on the 127-qubit \textit{ibm\texttt{\_}\!washington} device. GHZ state preparation circuits are constructed using the embedding algorithm. The algorithm selects qubit 73 as the source qubit, although this choice varied across calibration cycles due to physical error drift. Notably, all GHZ sizes are embedded with an optimal CNOT depth of 8, which is not possible on any of IBM Quantum's smaller heavy-hex devices due to boundary effects. We conduct five sets of experiments to obtain five independent measurements of the GHZ fidelity for each $N$. Each experiment requires $2N + 2$ circuits since the MQC method measures the overlap signal $S_\phi$ for $2N + 1$ values of $\phi$ (plus one circuit to measure the population). Prior to any GHZ experiments, we perform M3 readout error calibration. All circuits are executed with 4196 shots each.

\begin{figure}
    \includegraphics[width=.95\columnwidth]{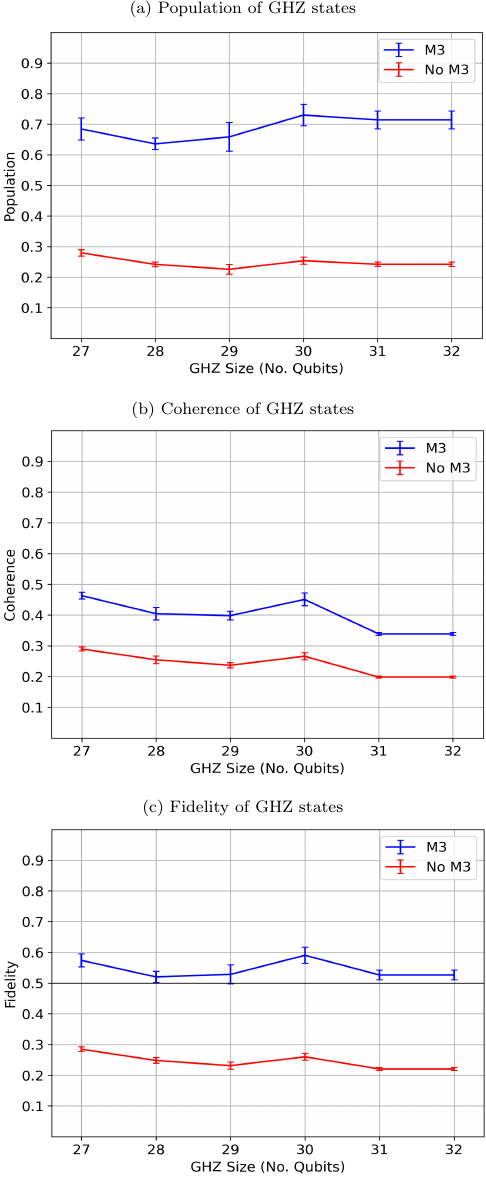}
    \caption{Measured populations \textbf{(a)}, coherences \textbf{(b)}, and fidelities \textbf{(c)} for GHZ states of size $N=27,28,29,30,31,32$ on the \textit{ibm\texttt{\_}\!washington} device. Plotted data points represent the mean value across five experiments, and error bars represent the standard error. We show results with and without readout error mitigation via M3. Showing an $N$-qubit GHZ state fidelity of at least 0.5 is sufficient to prove $N$-qubit GME. For the 32-qubit GHZ state, we calculate an unmitigated fidelity of $0.219\pm 0.006$ and a mitigated fidelity of $0.519\pm 0.014$.\label{fig:fidelities}}
\end{figure}

\begin{figure*}[t]
    \centering
    \includegraphics[width=\textwidth]{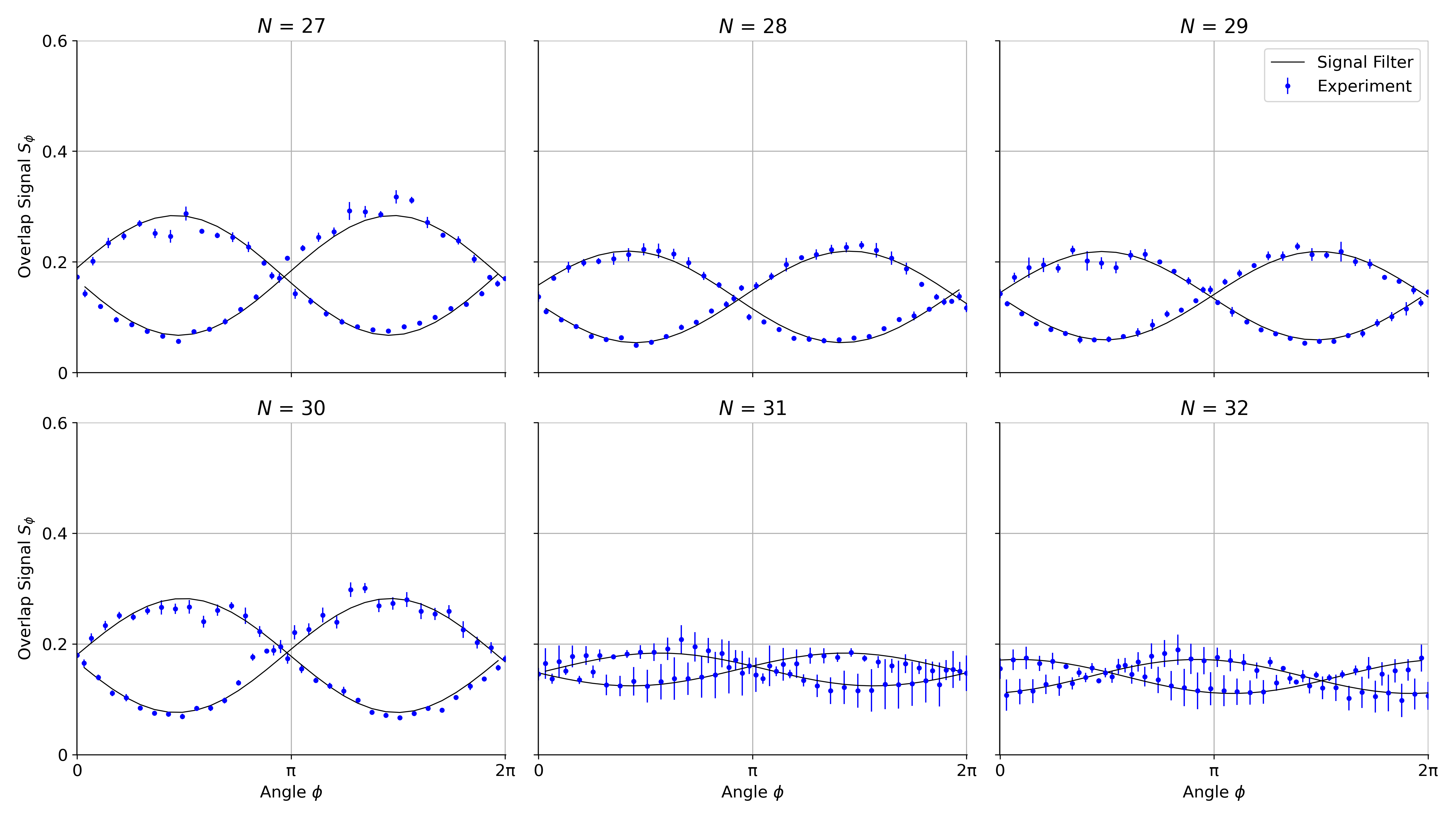}
    \caption{Measured MQC overlap signals for various size GHZ states on \textit{ibm\texttt{\_}\!washington}. Plotted values and error bars represent the mean and standard error across five independent experiments, respectively. All displayed results incorporate QREM via the M3 protocol. We plot a fitted signal curve for visualization purposes, however the actual signal amplitudes are filtered and calculated via fast fourier transform as in \cref{eq:signal amplitude}.}
    \label{fig:overlapsignal}
\end{figure*}

\Cref{fig:fidelities} shows population, coherence, and fidelity plots for GHZ states of size $N=27, 28, 29, 30, 31, 32$. Plotted data points display the mean value across five experiments and error bars represent the standard error. We plot results with and without readout error mitigation using M3. After applying readout correction, we measure the fidelity lower bounds for all states to be above the 0.5 threshold required to demonstrate GME. In particular, we measure a mitigated fidelity of $0.519\pm 0.014$ for the 32-qubit GHZ state. To the best of our knowledge, this is the largest GHZ state observed to have a fidelity of over 0.5. Furthermore, \cref{fig:overlapsignal} shows the corresponding MQC overlap signals (with QREM applied). The signal amplitudes are filtered and calculated using a fast fourier transform algorithm. The observed phase shift for certain $N$ is likely caused by free rotations in idle qubits.

The results are unusual in some aspects. Most obvious are the anomalously high fidelities for the 30-qubit GHZ state, where we measure a mitigated fidelity of $0.590 \pm 0.012$. Additionally, the measured populations seem to increase for GHZ states of size $N=30, 31, 32$, especially after M3 is applied. These peculiarities are likely due to several factors. Firstly, the experiments were executed in reverse order where the 32-qubit GHZ experiments were performed first. Due to the nature of the IBM Quantum job queuing system, not all experiments could be executed consecutively. In fact, experiments for GHZ size $N=30, 31, 32$ were performed in a different calibration cycle than experiments for GHZ size $N=27, 28, 29$. This is important because over the course of our research, we observed considerable performance drift in the \textit{ibm\texttt{\_}\!washington} device. In some attempts, the measured GHZ fidelities were well below the 0.5 threshold with no changes to the experimental procedure. We suspect that the discontinuity in the results reflects a decline in device performance during the latter half of experiments.

We draw comparisons to previous GHZ experiments on IBM Quantum devices \cite{Wei2020VerifyingCoherences,Mooney2021GenerationComputer,Yang2022EfficientDevices}. We observe a relatively small decrease in measured fidelities with increasing size. This may be explained by all GHZ sizes being prepared by the same CNOT depth of 8, in addition to the previously mentioned performance drift. We remark that the measured fidelities for similar size GHZ states on \textit{ibm\texttt{\_}\!washington} are not substantially higher than the GHZ fidelities on smaller devices from previous experiments \cite{Mooney2021GenerationComputer,Yang2022EfficientDevices} (in some cases being lower). This is not unexpected, since the average device error rates are often lower on the largest devices. We postulate that the larger verifiable GHZ state sizes on the 127-qubit device are partially enabled by its sheer scale---which allows larger GHZ states to be prepared with lower circuit depth. This highlights the importance of the scale of a quantum device in addition to the quality of its qubits. Next, we study the decay of GHZ states over time.

\section{Preserving GHZ States via Dynamical Decoupling}\label{section:dd}
\subsection{Hahn Echo and Periodic Dynamical Decoupling for GHZ States}\label{subsection:ghzpdd}
\begin{figure*}
    \includegraphics[width=.8\textwidth]{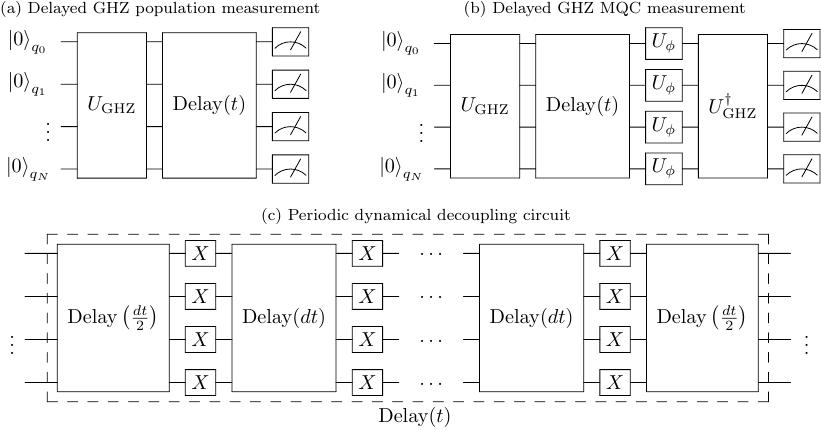}

    \caption{Circuit diagrams for delayed measurement of GHZ state fidelities \textbf{(a), (b)} and periodic dynamical decoupling \textbf{(c)}. The delays between control pulses of length $dt$ and the $X$-gate durations $t_X$ must sum up to the total delay $t$. For the special case of a Hahn echo (single control-pulse), we apply the $X$-gate in the middle of the delay period.\label{fig:circuitdelay}}
\end{figure*}

In this section, we explore the potential of dynamical decoupling techniques in prolonging the lifetimes of GHZ states on IBM Quantum devices. Dynamical decoupling is an open-loop control technique that mitigates decoherence in quantum computers by implementing sequences of control pulses \cite{Viola1998DynamicalSystems,Viola1999DynamicalSystems}. In theory, these control sequences effectively average out undesirable couplings between qubits and their environment. Dynamical decoupling can be seen as a generalization of the Hahn spin echo \cite{Hahn1950SpinEchoes}, which is the special case for a single or pair of control pulses. 

There are variations between dynamical decoupling schemes. The most basic scheme, known as periodic dynamical decoupling (PDD), applies equally spaced control pulses in quick succession. More advanced schemes include bounded-strength continuous sequences \cite{Viola2002RobustControls}, concatenated dynamical decoupling (CDD) \cite{Khodjasteh2005Fault-tolerantDecoupling}, and Uhrig dynamical decoupling (UDD) \cite{Uhrig2007KeepingSequences}. Different dynamical decoupling schemes may optimize for different scenarios and noise environments. There is ongoing research investigating the best way to integrate dynamical decoupling protocols with quantum computing algorithms. 

A typical dynamical decoupling technique on IBM Quantum devices is to implement control sequences during qubit idle periods. Most relevantly, dynamical decoupling has been shown to preserve GHZ coherence times by orders of magnitude on ion-trap qubits for up to four-qubit GHZ states \cite{Kaufmann2017ScalableEntanglement}. We study the efficacy of integrating dynamical decoupling with MQC in order to preserve GHZ fidelities. The experimental procedure follows from previous sections with the exception of adding a variable delay period between state preparation and fidelity measurements. We compare the decay of GHZ populations, coherences, and fidelities for free decays (idle qubits), and decays preserved with Hahn echo and PDD. The relevant circuit diagrams are shown in \cref{fig:circuitdelay}.

Decay experiments are performed for 7-qubit GHZ states prepared on the \textit{ibmq\texttt{\_}\!mumbai} device consisting of 27 qubits and a quantum volume \cite{Cross2019ValidatingCircuits} of 128. For free decays, we increment the circuit delay $t$ by 1 \textmu s up to a maximum delay of 15 \textmu s. For decays preserved with Hahn echo or PDD, we increment $t$ by 2 \textmu s up to a maximum of 30 \textmu s. For PDD, we implement control-pulses in 0.5 \textmu s regular intervals. We obtain both unmitigated results and results with QREM-applied via the M3 protocol. The GHZ state populations, coherences and fidelities are measured across five independent experiments, where circuits are executed with 4196 shots each. 

\begin{figure}[!ht]
    \centering
    \includegraphics[width=0.95\columnwidth]{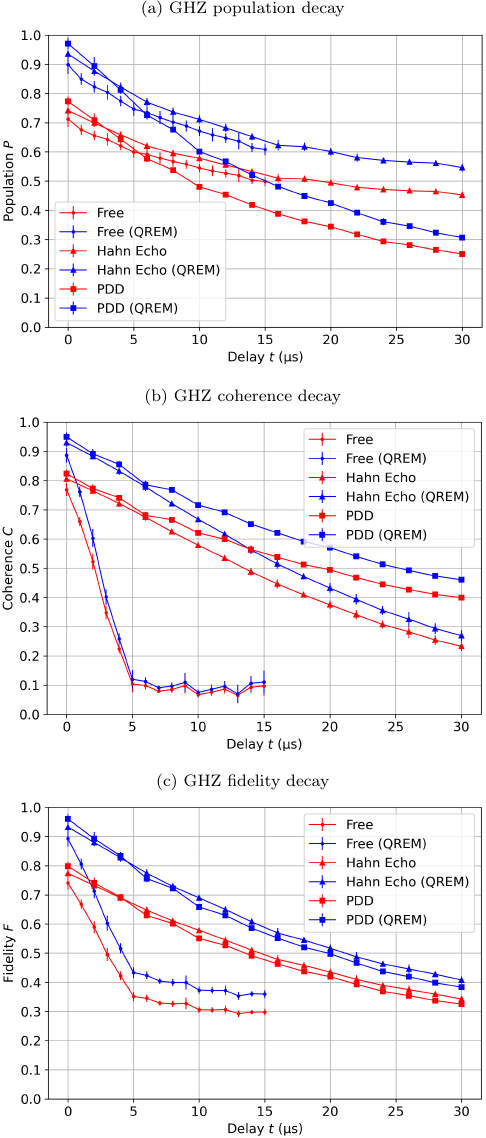}
    \caption{Population \textbf{(a)}, coherence \textbf{(b)}, and fidelity \textbf{(c)} as a function of circuit delay $t$ for 7-qubit GHZ states prepared on the \textit{ibmq\texttt{\_}\!mumbai} device. The coherences are measured using MQC. We compare the decay of the states without mitigation to the states preserved using Hahn echo (a single $\pi$-pulse) and periodic dynamical decoupling (PDD) with control-pulses applied every 0.5 \textmu s. We also display result before and after applying readout error mitigation.\label{fig:GHZdecays}}
\end{figure}

\Cref{fig:GHZdecays} displays the experimental results. As shown in \cref{fig:GHZdecays}(a), neither Hahn echo nor PDD led to a marked improvement of GHZ population times. In fact, PDD appears to accelerate the decay of GHZ populations. The results are related to how the ground and excited state populations evolve with respect to relaxation errors on superconducting quantum devices. In detail, qubits in the excited $\ket{1}$ state will eventually spontaneously decay into the ground $\ket{0}$ state at a rate described by the $T_1$ relaxation time. After a sufficient amount of time, a quantum computer will reset to the all ground $\ket{00\dots0}$ state (with some fluctuations). The decay in GHZ populations is primarily caused by bit flips in the $\ket{11\dots1}$ state due to thermal relaxation, although environmental noise can also cause random bitflips. The application of control pulses in Hahn echo or PDD, which flip the ground and excited state probabilities, will do little to prevent relaxation errors. In fact, as shown in the PDD curve, repeated application of $X$-gates only introduces additional noise from single-qubit gate errors.

\Cref{fig:GHZdecays}(b) shows substantial improvement in GHZ coherence times in experiments with Hahn echo and PDD. The GHZ coherences quantify the non-classical correlations between the $\ket{00\dots0}$ and $\ket{11\dots1}$ states. For free decays, this correlation drops to approximately 0.1 by $t=5$ \textmu s. In contrast, the PDD curve maintains a measured GHZ coherence of $C>0.4$ after $t=30$ \textmu s. Hahn echo also appreciably prolongs GHZ coherence times, resulting in $C>0.2$ at $t=30$ \textmu s, albeit to a lesser extent. These results highlight the efficacy of dynamical decoupling-based techniques in protecting GHZ states against dephasing errors (related to $T_2$ dephasing times) on superconducting quantum computers.

The decay of GHZ state populations and coherences are combined in \cref{fig:GHZdecays}(c), which plots the GHZ state fidelities as a function of circuit delay $t$. Both Hahn echo and PDD are shown to be effective techniques for preserving GHZ fidelities. It is interesting to observe that although PDD is superior to Hahn echo for preserving GHZ coherences, it is worse at preserving GHZ populations. This is likely due to additional noise introduced from the PDD gate sequences, which has the effect of obfuscating population measurements in exchange for better protection against decoherence. As a result, Hahn echo exhibits roughly similar performance to PDD in preserving overall fidelities. In future experiments, it may be worth testing more advanced dynamical decoupling protocols such as UDD \cite{Uhrig2007KeepingSequences}, which are shown to be more typically more effective than PDD \cite{Ezzell2022DynamicalSurvey}. We comment that the small differences in the initial values of $P$, $C$ and $F$ are likely attributed to device drift. We also remark that applying QREM mainly increases the initial $P$, $C$ and $F$ values with little influence to the decay rates. Next, we evaluate the scaling of GHZ decoherence rates as a function of state size.

\subsection{Scaling of Decoherence Rates and GHZ Size }\label{subsection:ratescaling}
Studying the decoherence rates of multipartite states as a function of their size may provide vital insight into the noise underlying a quantum system. The strength and nature of this noise can determine the feasibility of scaling up a quantum device. In particular, it may reveal whether a system exhibits superdecoherence. Superdecoherence describes the coupling of qubits to a single reservoir, which cause qubit decoherence rates to scale with the size of the system \cite{Kattemolle2020ConditionsSuperdecoherence}. Such an effect is detrimental to the realization of large-scale, fault-tolerant quantum computers. GHZ states are particularly convenient for detecting superdecoherence due to their high sensitivity to noise. In detail, GHZ states accumulate decoherence between qubits, so if the dominant noise model is uncorrelated across qubits (i.e. the decoherence rate per qubit is constant), we expect GHZ decoherence rates to scale linearly with the number of qubits. In contrast, if the dominant noise model is correlated across qubits (i.e. the system experiences superdecoherence), we expect GHZ decoherence rates to scale polynomially.

\begin{figure*}
    \centering
    \includegraphics[width=0.65\textwidth]{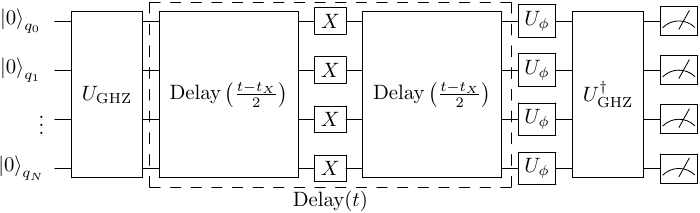}
    \caption{Circuit for measuring GHZ decoherence rates as a function of the number of qubits, as in \cref{fig:circuitdelay}. We set a delay of $t$ between GHZ state preparation and MQC measurement. We implement a single $\pi$-pulse in the middle of the delay period to extend GHZ coherence times.\label{fig:scalingcirc}}
\end{figure*}

GHZ decoherence rates as a function of state size have been studied on ion-trap and superconducting quantum devices for up to state size 6 and 8, respectively \cite{Monz201114-qubitCoherence,Ozaeta2019DecoherenceProcessor}. Prominently, the ion-trap device was shown to exhibit quadratic scaling of GHZ decoherence rates, indicating superdecoherence (note that this does not imply the same for all ion-trap systems). On the other hand, the IBM Quantum superconducting device displayed linear GHZ scaling. In this section, we extend the study on GHZ decoherence scaling on IBM's more recent superconducting devices, for GHZ states of up to 15 qubits in size. Furthermore, for the first time, we incorporate readout error mitigation and measure the coherences via MQC, incorporating Hahn echo.

To provide easy comparison, we conduct our experiments in a manner similar to Ozaeta \& McMahon's previous study involving IBM Quantum devices \cite{Ozaeta2019DecoherenceProcessor}. The study, undertaken in 2018, measured GHZ decay rates on the now retired 16-qubit \textit{ibmq\texttt{\_}\!melbourne} device, which employed a square lattice qubit topology. Today, we implement our study on the 27-qubit \textit{ibm\texttt{\_}\!hanoi} device, which employs a heavy-hex topology. In contrast to Ozaeta \& McMahon's study which measures GHZ coherences using parity oscillations, we employ MQC, incorporating dynamical decoupling-based techniques in the form a single $\pi$-pulse.

The circuit for measuring GHZ decoherence rates is shown in \cref{fig:scalingcirc}. To prolong coherence times, we implement a single $\pi$-pulse in the middle of the delay period. For an $N$-qubit GHZ state, we model the coherence $C^{(N)}$ as a function of delay $t$ as the exponential decay
\begin{equation}\label{eq:decoherence}
    C^{(N)}(t) = C_0^{(N)}e^{-\alpha^{(N)}t}
\end{equation}
where $C_0^{(N)}=C^{(N)}(t=0)$ is the initial coherence, and $\alpha^{(N)}= 1/T_{GHZ}^{(N)}$ is the decoherence rate, where its reciprocal $T_{GHZ}^{(N)}$ is the GHZ coherence time.

\begin{figure*}
    \centering
    \includegraphics[width=\textwidth]{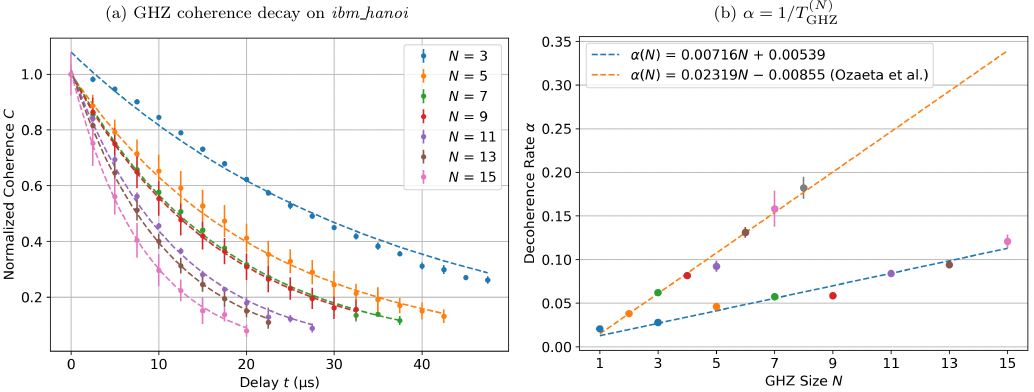}
    \caption{\textbf{(a)} GHZ normalized coherences for various state sizes as a function of circuit delay $t$. Data points represent the coherence averaged over five experiments. Error bars represent the standard deviation. The dashed lines graph the fitted exponential decay curve from \cref{eq:decoherence}. \textbf{(b)} The GHZ decoherence rate $\alpha = 1/T_{\mathrm{GHZ}}^{(N)}$ as a function of state size $N$. The error bars represent the standard deviations obtained from the covariance matrix produced by the curve fitting algorithm. We fit a linear trend over the plotted $alpha$ values, drawn by the dashed line. The straight line equation is given by $\alpha (N) = (7.16N + 5.39) \cdot 10^{-3}$ \textmu s$^{-1}$ with $R^2 = 0.962$. We include results from Ozaeta et al. \cite{Ozaeta2019DecoherenceProcessor} for comparison.\label{fig:coherencedecays}}
\end{figure*}

\Cref{fig:coherencedecays}(a) plots the $N$-qubit GHZ state coherences (normalized) as a function of circuit delay $t$ for $N=3,5,7,9,11,13,15$ on $\textit{ibm\texttt{\_}\!hanoi}$, which consists of 27 qubits and a quantum volume of 64. We increase total delay $t$ in increments of 2.5 \textmu s. The maximum delay for each experiment ranges from $t_{\mathrm{max}}=47.5$ \textmu s for $N=3$ to $t_{\mathrm{max}}=20.0$ \textmu s for $N=15$. For reference, the average CNOT gate time on \textit{ibm\texttt{\_}\!hanoi} is 385 ns. Data points represent the average measured coherence between five experiments, and error bars represent the standard deviation. We fit the plotted data with the exponential decay curve \cref{eq:decoherence}. We execute circuits with 4196 shots each. We mitigate readout errors using M3, although as shown in the previous section, this has little effect on the decay rates. 

Most of the decay curves are modeled well by the exponential decay function. A notable exception, however, is the $N=3$ coherences, which appear to plateau slightly before exhibiting exponential decay. As a result, the decay fit incorrectly projects the initial coherence $C^{(N=3)}_0$ to be greater than 1. We observe a clear pattern of increasing decay rates with increasing GHZ size. In order to quantify this trend, we plot the decoherence rate $\alpha$ and the GHZ size $N$, shown in \cref{fig:coherencedecays}(b). We take the error from the standard deviation, which we derive from the covariance matrix produced by the fitting algorithm. In agreement with Ozaeta \& McMahon, we are able to well fit the data with the linear trendline $\alpha (N) = (7.16N + 5.39)10^{-3}$ \textmu s$^{-1}$ with $R^2 = 0.962$---now extending up to $N=15$. (With an anomoalous data point at $N=9$). Our results support the notion that recent IBM Quantum transmon devices are naturally robust against superdecoherence. We remark that although we only measure coherence times for GHZ states of up to 15 qubits, MQC and dynamical decoupling-based techniques improve the initial $C_0$ values enough that one can feasibly extend the study to larger GHZ states, especially as error rates and coherence times improve.

\begin{table}
    \normalsize
    \centering
    %\rowcolors{2}{white}{gray!15}
    \renewcommand{\arraystretch}{1.2}
    \begin{tabular}{ c c c c }
        \hline\hline
        $N$ & $\alpha$ [$\textrm{\textmu s}^{-1}$] (MQC) & $T_{\mathrm{GHZ}}^{(N)}$ [\textmu s] (MQC) & $T_{\mathrm{GHZ}}^{(N)}$ [\textmu s] \cite{Ozaeta2019DecoherenceProcessor}\\
        \hline
        1 & - & - & $48.34\pm 1.56$\\
        2 & - & - & $26.15\pm 1.67$\\
        3 & $(27.96 \pm 0.27) 10^{-3}$ & $35.77 \pm 0.35$ & $16.11\pm 0.89$\\
        4 & - & - & $12.25\pm 0.62$\\
        5 & $(45.84 \pm 1.46) 10^{-3}$ & $21.81 \pm 0.70$ & $10.83\pm 0.75$\\
        6 & - & - & $7.63\pm 0.36$\\
        7 & $(57.50 \pm 1.36) 10^{-3}$ & $17.39 \pm 0.41$ & $6.32\pm 0.83$\\
        8 & - & - & $5.49\pm 0.38$\\
        9 & $(58.80 \pm 3.00) 10^{-3}$ & $17.01 \pm 0.87$ & -\\
        11 & $(83.66 \pm 1.03) 10^{-3}$ & $11.95 \pm 0.15$ & -\\
        13 & $(93.95 \pm 3.59) 10^{-3}$ & $10.64 \pm 0.41$ & -\\
        15 & $(120.27 \pm 8.01) 10^{-3}$ & $8.31 \pm 0.55$ & -\\
        \hline\hline
    \end{tabular}
    \caption{Fitted values of GHZ decoherence rates $\alpha$ and coherence times $T_\mathrm{GHZ}$ where $T_\mathrm{GHZ} = 1/\alpha$. For comparison, we include results from Ozaeta \textit{et al}. \cite{Ozaeta2019DecoherenceProcessor} from experiments on previous IBM Quantum devices.}
    \label{tab:coherencetable}
\end{table}

We summarize the GHZ coherence times on \textit{ibm\texttt{\_}\!hanoi} in \cref{tab:coherencetable}. For easy comparison, we include the measured GHZ coherence times from Ozaeta \& McMahon's experiments. For $N=3,5,7$, we report an average increase of coherence times of 133\% over previous results. Furthermore, we report longer coherence times on 15-qubit GHZ states than on previous 8-qubit GHZ states of $8.31 \pm 0.55$ μs and $5.49 \pm 0.38$ μs, respectively. The improvement in GHZ coherence times may be attributed to hard improvements in combination with superior coherence detection methods, incorporating dynamical-decoupling based techniques. %We comment that it is possible to extend GHZ coherence times even further by implementing full dynamical decoupling as demonstrated in \ref{section:dd}.

\section{Efficient Bipartite Entanglement Characterization in Whole-Device Graph States}\label{section:graphstate}
\subsection{Graph States}
Graph states are a class of entangled multi-qubit states that are defined with respect to a connected graph. They are a generalization of cluster states \cite{Briegel2001PersistentParticles}, and hence form a universal basis for measurement-based computation \cite{Raussendorf2001AComputer}. Graph states are additionally useful for quantum error correcting codes \cite{Schlingemann2002QuantumGraphs}, quantum secure communication \cite{Markham2008GraphSharing}, quantum metrology \cite{Shettell2020GraphMetrology}, and probing Bell inequalities over multi-partite systems \cite{Guhne2005BellStates}. In quantum circuit notation, a graph state may be expressed as
\begin{equation}
    \ket{G_n} = \prod_{(a, b) \in E} \mathrm{CZ}^a_b \ket{+}^{\otimes n}
\end{equation}
where \(\ket{+} = \frac{1}{\sqrt{2}}(\ket{0} + \ket{1})\), \(E\) is the set of edges connecting graph \(G_n\) containing \(n\) vertices (qubits), and \(\mathrm{CZ}^a_b\) represents a controlled-phase gate between adjacent qubits \(a\) and \(b\). We can equivalently define the \textit{stabilizing operator} \(S_a\) for each qubit \(a\) in \(G\):

\begin{equation}
    S_a = \sigma_x^{(a)} \prod_{b \in N(a)}\sigma_z^{(b)}
\end{equation}
where \(\sigma_{x,y,z}^{(a)}\) are Pauli operators acting on qubit \(a\) and \(N(a)\) is the set of qubits adjacent to \(a\). Thus \(\ket{G_n}\) is the simultaneous \(+1\) eigenstate of \(n\) operators following \(S_a\ket{G_n} = \ket{G_n}\).

\begin{figure}
    \centering
    \includegraphics[width=0.85\columnwidth]{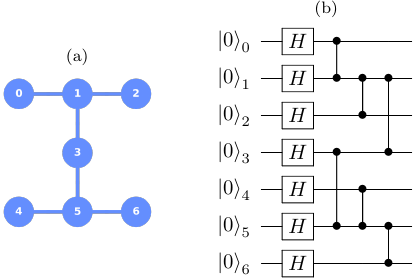}
    \caption{Native-graph state preparation circuit \textbf{(b)} for seven-qubit layout \textbf{(a)}. Firstly, prepare each qubit into an equal superposition by applying Hadamard gates on the $\ket{0}$ states. Next, apply a controlled-phase (CZ) corresponding to each edge on the device graph. The two-qubit gate depth of the preparation circuit is at least the maximum node degree of the represented graph.\label{fig:7qgraphstate}}
\end{figure}

Graph states are a convenient choice for studying large-scale entanglement as they are simple to prepare and comparitively noise robust \cite{Briegel2001PersistentParticles}. Controlled-phase operations that do not overlap vertices can be applied in parallel, allowing any graph state to be prepared by a linear-size constant-depth circuit \cite{Hyer2006ResourcesStates}. Concretely, a bounded degree graph can be prepared with a two-qubit gate depth equal to the maximum degree between its vertices \cite{Aliferis2004ComputationPicture}. An example least-depth graph state preparation circuit on a seven-qubit layout is shown in \cref{fig:7qgraphstate}. The qubit layout is represented in \cref{fig:7qgraphstate}(a) where nodes represent qubits and edges display possible CNOT operations. Since the graph has a maximum node degree of three, its graph state preparation circuit has a two-qubit gate depth of three.

To characterize bipartite entanglement across an entire device, we prepare a native-graph state containing every edge. We then perform full quantum state tomography on every locally entangled bipartition corresponding to each edge on the device. In detail, graph states have the property that projecting all but two qubits in an entangled cluster leaves the pair in a maximally entangled Bell state \cite{Raussendorf2003Measurement-basedStates}. The extent of two-qubit entanglement can then be quantified by measuring the negativity \cite{Vidal2002ComputableEntanglement}.  For a quantum state represented by the density matrix $\rho$, the negativity $\mathcal{N}(\rho)$ between subsystems $A$ and $B$ is calculated as
\begin{equation}\label{eq:negativity2}
    \mathcal{N}(\rho) = \left\lvert\sum_{\lambda_i<0}\lambda_i\right\rvert = \sum_i \frac{\lvert \lambda_i\rvert - \lambda_i}{2}
\end{equation}
where $\lambda_i$ are the eigenvalues of $\rho^{T_B}$, where $\rho^{T_B}$ is the partial transpose of $\rho$ with respect to subsystem $B$. A maximally entangled Bell state has a negativity of \(\mathcal{N} = 0.5\), whereas a fully separable state has a negativity of \(\mathcal{N} = 0\). Although there are many other entanglement measures, including more complex multipartite entanglement witnesses, negativity is an entanglement monotone that is simple to compute. A non-zero measurement for the negativity on a 2-qubit state is a necessary and sufficient condition for entanglement. This makes it a great choice for bipartite entanglement on graph states, where edges can be reduced to Bell states. Additionally, negativity is related to the minimum teleportation distance $d_\mathrm{min}(\rho)$ achievable with state $\rho$ acting on \(\mathbb{C}^m \otimes \mathbb{C}^m\):
\begin{equation}\label{eq:teleportation}
    d_\mathrm{min}(\rho) \geq \frac{2}{m+1}(m-1+2\mathcal{N}(\rho))
\end{equation}
A device is said to be whole-device entangled if every qubit is connected to the main graph where edges correspond to qubit pairs with a measured negativity of \(\mathcal{N} > 0\). This is distinct from saying the qubits are genuinely multipartite entangled (which follows a more strict criteria), but rather, there exist no bipartition of qubits on the device that results in separable states.

\subsection{Bipartite Entanglement Characterization Protocol}
We develop a protocol to efficiently characterize bipartite entanglement on quantum computers, inspired by experiments from Mooney \textit{et al}. \cite{Mooney2019EntanglementComputer,Mooney2021Whole-deviceComputer}. Development of the scheme is driven by three main design principles: to devise an entanglement characterization protocol that is highly automated, scalable, and architecture-independent. We implement the protocol in Python, utilizing the Qiskit API to interface with IBM superconducting quantum devices. However, the techniques and procedures in the program are generally applicable. The protocol can be divided into five components:
\begin{enumerate}
    \item Native-graph state preparation. Automatically construct a whole-device graph state preparation circuit which entangles every qubit on the device.
    \item Parallel Quantum State Tomography. Execute quantum state tomography circuits on Bell states prepared on the graph state in parallel. For heavy-hex qubit architectures, this step can be performed in four batches of nine circuits each.
    \item Quantum readout error mitigation. Mitigate readout errors using classical post-processing.
    \item Density matrix reconstruction. Reconstruct the Bell state density matrices using readout error mitigated or unmitigated measurement results.
    \item Negativity calculation and entanglement mapping. Calculate the bipartite negativities corresponding to each qubit pair on the device and construct the entanglement graph.

\end{enumerate}

The first and second component, which contain techniques unique to this work \cite{Mooney2019EntanglementComputer, Mooney2021Whole-deviceComputer}, are elaborated below.

\begin{figure*}
    \centering
    \includegraphics[width=0.66\textwidth]{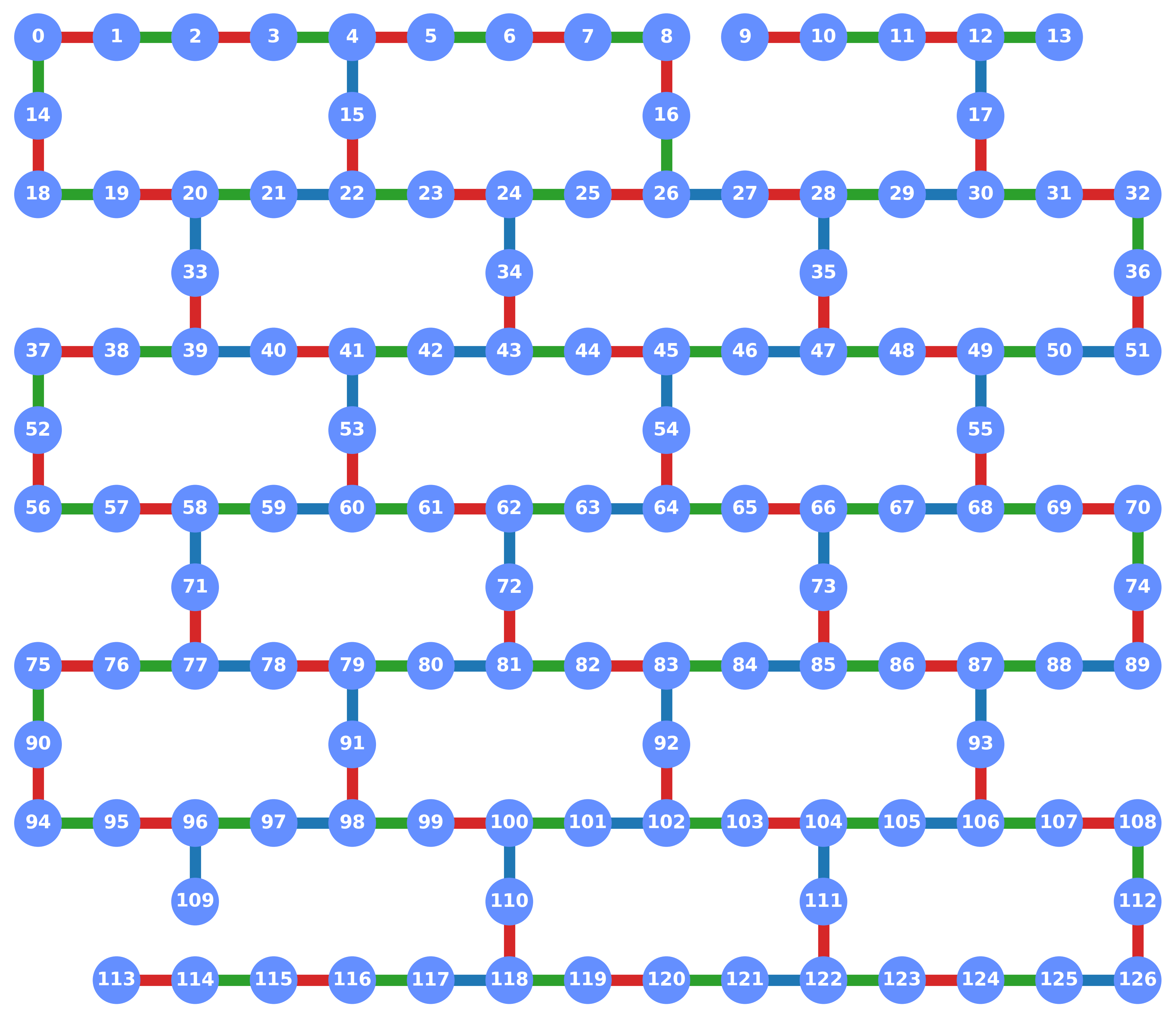}
    \caption{Graph state preparation on the Eagle r1 processor (127 qubits). Edges represent controlled phase gates between qubits, where red edges are applied at depth one, green edges at depth two, and blue edges at depth three. The device graph is defined by a heavy-hexagonal lattice (or subdivided honeycomb).}
    \label{fig:dmap127}
\end{figure*}

\subsubsection{Native-Graph State Preparation}
Similar to the GHZ case, the objective is to embed depth-optimal circuits using an automated routine. (In previous works \cite{Wang201816-qubitEntangled,Mooney2019EntanglementComputer,Mooney2021Whole-deviceComputer}, this is done manually) Since heavy-hex lattices have a maximum node degree of three, it is possible to embed a native-graph state circuit with a minimum two-qubit gate depth of three. Topology-specific methods for constructing optimal graph state circuits exist, such as stitching together smaller circuits embedded on unit cells \cite{Mooney2021Whole-deviceComputer}. It is sufficient, however, to implement a greedy algorithm that applies as many CNOT gates in parallel in each step. The algorithm also has the advantage of working with any qubit layout, although it is unknown if it is universally optimal. A whole-device graph state embedding prepared by the algorithm on the 127 qubit Eagle processor is shown in \cref{fig:dmap127}. Additional graph state embeddings on other IBM physical layouts are shown in the results section below.

\subsubsection{Parallel Quantum State Tomography}
In previous graph state experiments \cite{Mooney2019EntanglementComputer,Mooney2021Whole-deviceComputer}, full quantum state tomography (QST) is performed on one qubit pair (and its neighbours) at a time. Using such method, the number of circuits required to fully characterize a device increases linearly with the number of edges. Our procedure improves upon this by executing QST in parallel. In detail, we perform simultaneous basis measurements on non-overlapping sets of qubits, where each set defines a target qubit pair and its neighbours. By grouping sets into batches and performing parallel QST for batches at a time, we can fully characterize bipartite entanglement on any size device (provided invariable qubit topology) with a constant number of circuits. 

Using a specialized scheduling scheme for the heavy hex hardware layout, the number of required batches can be reduced to only four when allowing Bell pairs to share neighbours. This can be achieved by using an alternate tiling of two unit cells (able to be rotated 180\textdegree), where each unit cell is composed of six edges and each edge is assigned to one of four batches. However, similar to the case of graph state embedding, we instead opt for a topology-agnostic algorithm that performs parallel QST for as many non-overlapping Bell pairs as it can fit into a single batch. In our implementation, we prevent target pairs from sharing neighbours for practical convenience. \Cref{tab:batches} lists the number of batches required to characterize IBM Quantum devices up to 433 qubits in size.

\begin{table*}[t]
    \normalsize
    \centering
    \renewcommand{\arraystretch}{1.2}
    \begin{tabular}{l c c c}
        \hline\hline
        Device(s) & Qubits & Batches & Circuits\\ \hline
        \textit{manila, belem, lima, quito, santiago, bogota} & 5 & 4 & 36 \\
        \textit{perth, jakarta, lagos} & 7 & 6 & 54 \\ 
        \textit{guadalupe} & 16 & 6 & 54 \\ 
        \textit{kolkata, mumbai, toronto, montreal, hanoi, cairo} & 27 & 6 & 54 \\ 
        \textit{brooklyn, ithaca} & 65 & 6 & 54 \\ 
        \textit{washington, sherbrooke, brisbane} & 127 & 8 & 72 \\ 
        \textit{seattle} & 433 & 8 & 72 \\
        \hline\hline
        
    \end{tabular}
    \caption{Number of tomography batches and total circuits required by the batch-finding algorithm for full QST on IBM Quantum devices.}
    \label{tab:batches}
\end{table*}

The table shows that the batching algorithm collates tomography circuits into 6-8 batches for heavy-hex devices. Performing full QST on $n$ qubits requires $3^n$ circuits corresponding to each combination of Pauli bases. Therefore each batch, which performs two-qubit QST in parallel, contains nine circuits. The variation in the number of batches is likely attributed to the greedy nature of the algorithm, which may group sets of qubits into batches in non-optimal order. Nevertheless, the procedure's main utility lies in reducing the number of tomography circuits to a roughly constant number in addition to being compatible with various qubit topologies.

\subsection{Bipartite Entanglement on IBM Quantum Devices}
We characterize bipartite entanglement on all IBM Quantum devices accessible by the University of Melbourne IBM Quantum Hub. At the time of experiment, these include four 5-qubit systems, five 7-qubit systems, one 16-qubit system, eight 27-qubit systems, three 127-qubit systems, and one 433-qubit system, totaling to 22 systems. Both unmitigated and mitigated results using QREM are shown. It is important to present unmitigated results because not all protocols involving graph states can incorporate readout error mitigation. Notably, quantum teleportation schemes which use mid-circuit measurements are incompatible with QREM. 

We perform eight sets of graph state experiments per device, sampling all circuits with 8192 shots. Readout error calibration circuits are sampled with the same number of shots. All native-graph states are prepared with the optimal two-qubit gate circuit depth of three. The number of circuits per experiment for various size devices is shown in \cref{tab:batches}. Besides practicality, reducing the number of circuits per experiment is beneficial because it also reduces the variability in results due to device drift. To assign a single negativity for each device edge, we calculate the mean \textit{maximum} negativity between experiments, where maximum refers to the largest negativity between possible Bell state projections. We take the error to be standard error.

\begin{figure*}
    \centering
    \includegraphics[width=\textwidth]{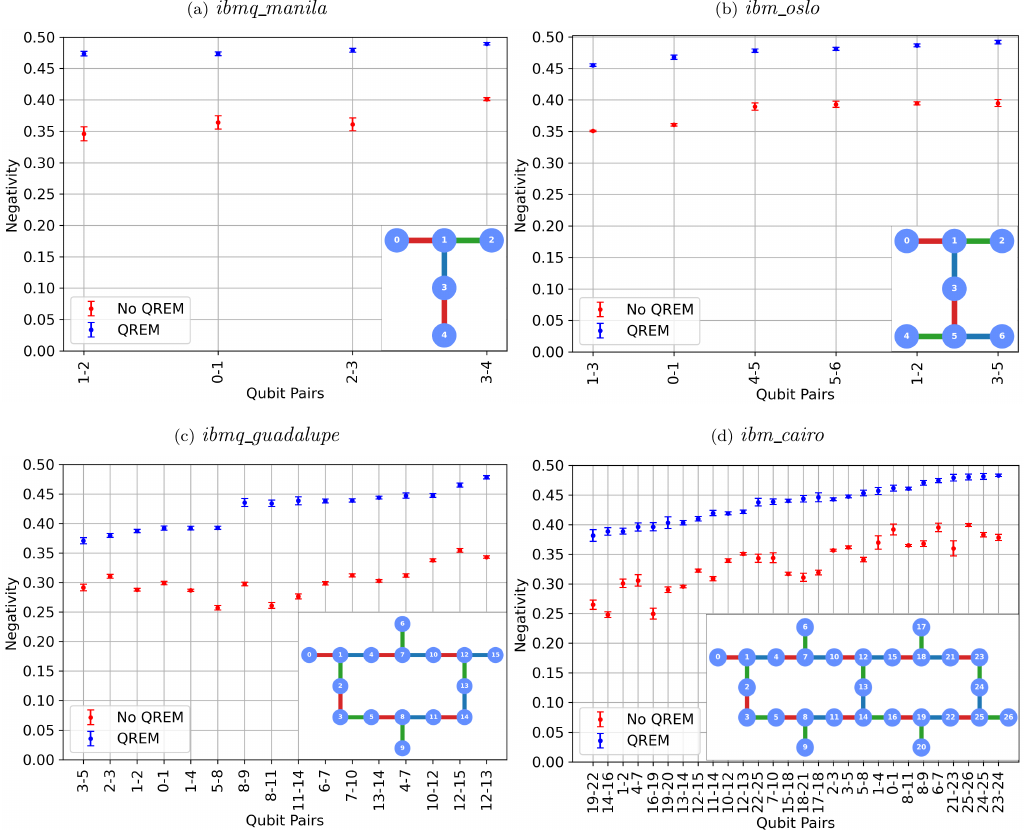}
    \caption{Negativity plots of four various-size devices and their graph state embeddings. The edge negativity is taken to be the maximum negativity between Bell state projections on neighbouring qubits, averaged over eight experiments. A negativity of 0 represents no entanglement, whereas a negativity of 0.5 represents maximal entanglement. The error bars show the standard error. The mean unmitigated device negativities for ibmq\texttt{\_}\!manila, ibm\texttt{\_}\!oslo, ibmq\texttt{\_}\!guadalupe, and ibm\texttt{\_}\!cairo are 0.388, 0.403, 0.335, and 0.356 respectively. After quantum readout error mitigation (QREM), the mean device negativities are 0.487, 0.488, 0.447, and 0.455 respectively, representing an average percent improvement of 26.3\%. All four systems exhibit whole-device entanglement. The full data table summarizing all results\label{fig:negativities}}
\end{figure*}

\Cref{fig:negativities} show sample negativity plots for devices up to 27 qubits in size and their respective graph state embeddings. Edges are sorted in order of ascending lower bounds of mitigated negativities. Among these systems, \textit{ibm\texttt{\_}\!oslo} purports both the highest mitigated and unmitigated mean device negativities of 0.488 and 0.403, respectively. Nevertheless, all negativity plots indicate that each of these systems exhibit whole-device entanglement. The improvement in negativities due to QREM is consistently significant. The average percentage improvement in mean negativity across the four devices is 26.3\%

\begin{figure*}
    \centering
    \includegraphics[width=\textwidth]{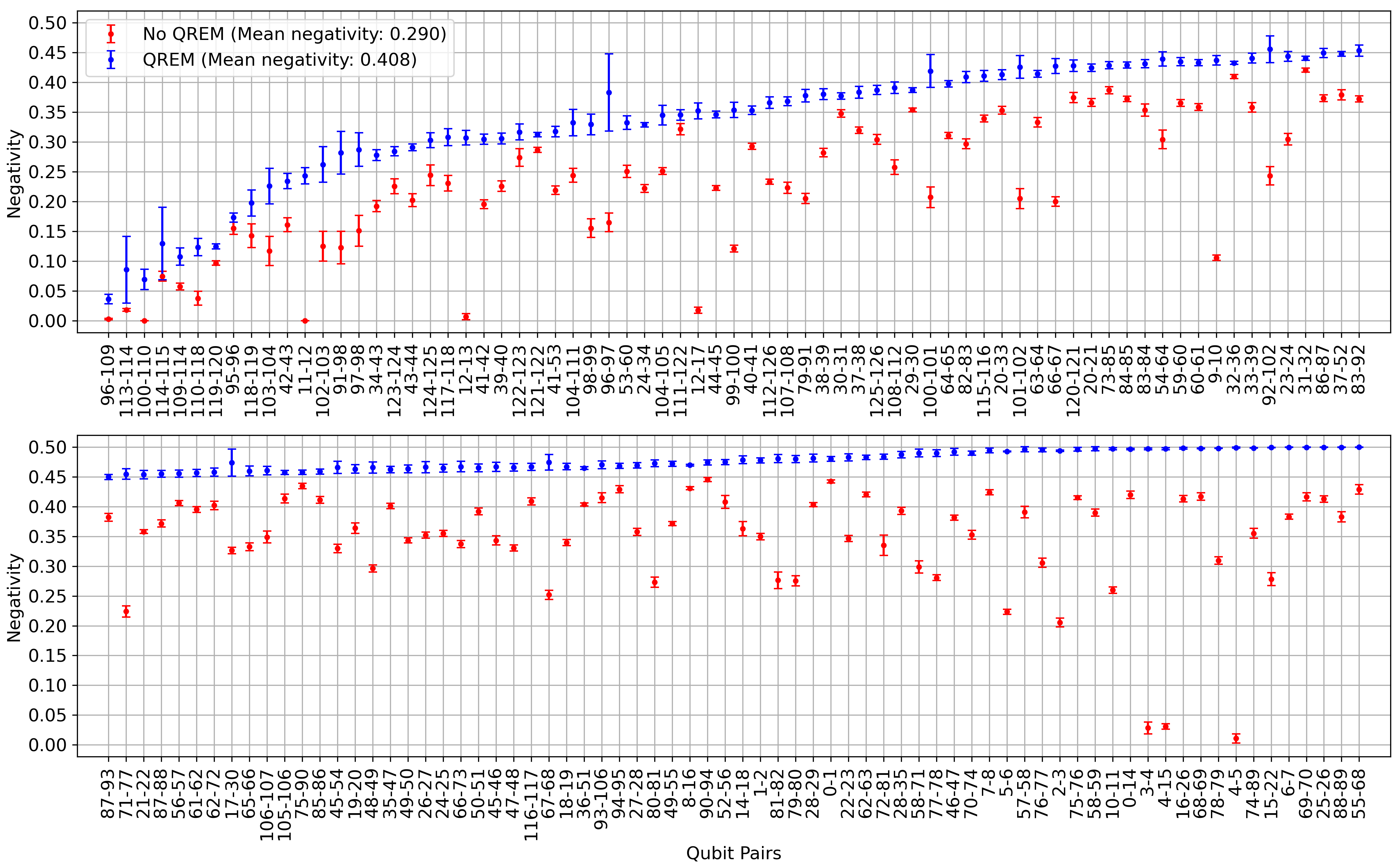}
    \caption{Negativity plot for 127-qubit device \textit{ibm\texttt{\_}\!washington}. The native-graph state circuit embedding is shown in \cref{fig:dmap127}. The unmitigated negativities have a mean of 0.290 and a standard deviation of 0.117. The mitigated negativities have a mean of 0.408 and a standard deviation of 0.102. Qubits 4 and 12 display abnormally high readout error rates of 0.338 and 0.390 respectively. These correspond to the large gaps between mitigated and unmitigated negativity values for edges 4--5, 4--15, 3--4, 12--17, 12--13, and 11--12.}
    \label{fig:negativities2}
\end{figure*}

\Cref{fig:negativities2} shows the negativity plot for 127-qubit device \textit{ibm\texttt{\_}\!washington}, with the respective graph state embedding is shown in \cref{fig:dmap127}. The unmitigated negativities have a mean of 0.290 and a standard deviation of 0.117. After implementing QREM, the resulting negativities have a mean of 0.408 and a standard deviation of 0.102---a 40.7\% percent improvement in mean device negativity. We comment on several anomalous edges with large gaps between the mitigated and unmitigated negativity, such as edges 4--5, 4--15. These can be attributed to significantly higher than average readout error rates for certain qubits. In particular, qubits 4 and 12 display abnormally high readout error rates of 0.338 and 0.390 respectively, corresponding to the large negativity gaps at edges 4--5, 4--15, 3--4, 12--17, 12--13, and 11-12.

\begin{figure*}
    \centering
    \includegraphics[width=\textwidth]{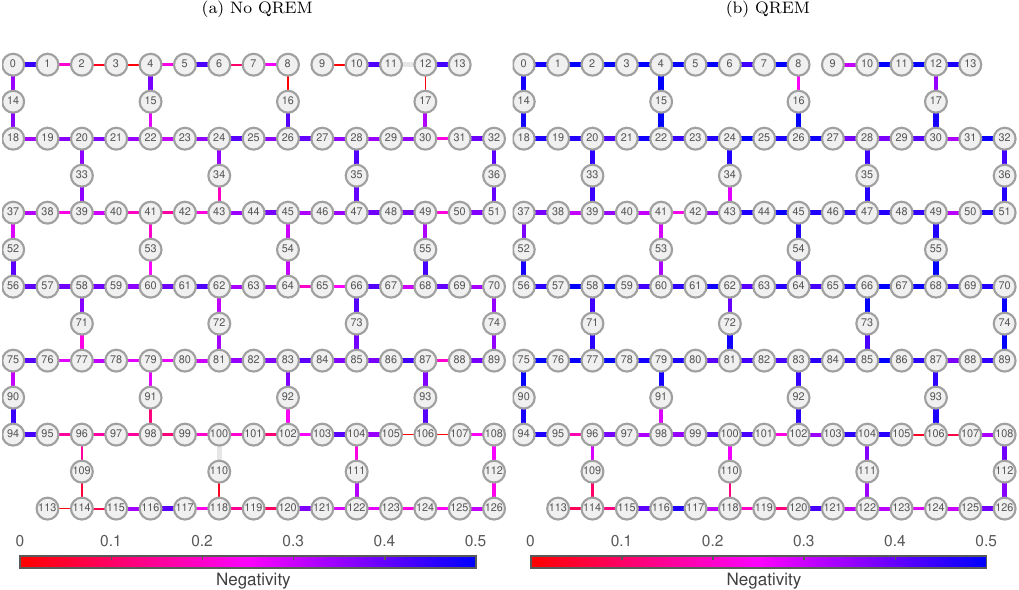}
    \caption{A graphic representation of entanglement within native-graph states prepared on 127-qubit device \textit{ibm\texttt{\_}\!washington}. Both unmitigated results \textbf{(a)} and results mitigated with QREM \textbf{(b)} are shown. Thin red and thick blue edges represent minimal and maximal entanglement, respectively. Qubits not connected (entangled) to the main graph are greyed out. In (a), there are 6 qubits disconnected from the main graph. In (b), we observe whole-device entanglement. Notably, pairs with low entanglement tend to concentrate in regions, such as in the lower left corner.}
    \label{fig:negativitymap}
\end{figure*}

As IBM Quantum's earliest Eagle processor, \textit{ibm\texttt{\_}\!washington} purports a lower mean device negativity than most 27-qubit Falcon processors and the newer Eagle r3 devices (see \cref{tab:summary}). Although, after applying QREM, we observe whole-device entanglement across all 127 qubits. To illustrate, \cref{fig:negativitymap} draws graphic representations of entanglement within \textit{ibm\texttt{\_}\!washington}. Negativity values are mapped on device edges where thin red edges represent low negativity and thick blue edges represent high negativity. Edges with lower-bound negativities of zero are greyed out. We remark that edges with low negativity tend to coalesce in regions. These areas of low entanglement, such as in the lower left corner of \cref{fig:negativitymap}, may arise due to physical factors such as non-uniform heat distribution in the device.

\begin{figure*}
    \includegraphics[width=\textwidth]{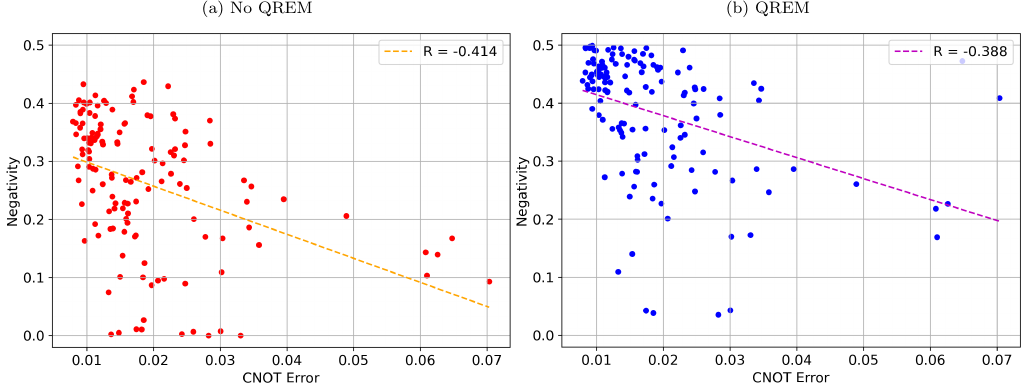}
    \caption{Negativity versus CNOT error rates on 127-qubit \textit{ibm\texttt{\_}\!washington} with and without QREM. The CNOT error is taken to be the average CNOT error between the Bell state pair in addition to its neighbouring qubits. Figure \textbf{(a)} shows a Pearson correlation of $R=-0.414$ between unmitigated negativities and CNOT error rates, whereas figure \textbf{(b)} shows a correlation of $R=-0.388$ between mitigated negativities and CNOT error rates. These values lie within general expectation, since CNOT errors should negatively correlate with entanglement measures.\label{fig:negativitycorrelation}}
%\end{figure*}
%\begin{figure*}
    \includegraphics[width=\textwidth]{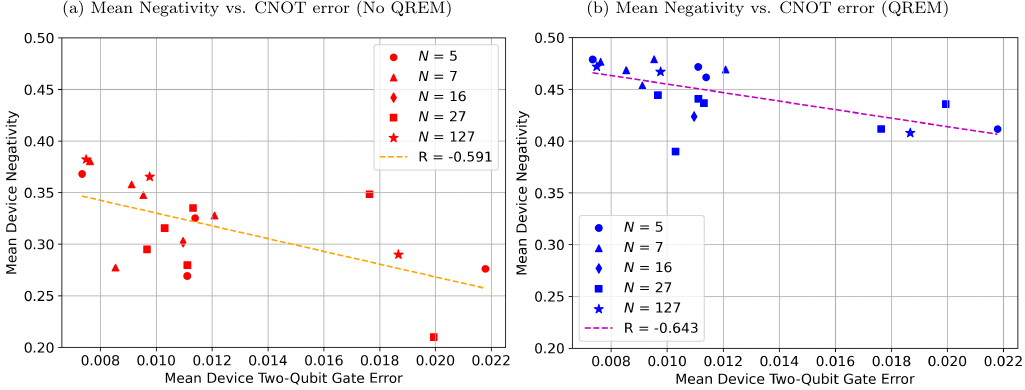}
    \caption{Mean device negativity versus mean device CNOT error rate for various device size $N$. We take the device negativities to be the maximum negativities between bins. \textbf{(a)} plots the unmitigated negativities and \textbf{(b)} plots the negativities with QREM applied. We observe a correlation of $R=-0.591$ and $R=-0.634$ for unmitigated and readout error mitigated results, respectively.\label{fig:devicecorrelation}}
\end{figure*}

To further our investigation of entanglement within the 127-qubit device, we plot the Bell state negativities against CNOT error rates in \cref{fig:negativitycorrelation}. Precisely, we take the CNOT error to be the average CNOT error between edges in the tomography set, which in addition to the Bell state pair, includes its adjacent neighbours. Furthermore, unlike previous figures, we take the negativity as the mean between projections on adjacent qubits rather than the maximum. \Cref{fig:negativitycorrelation} shows Pearson correlation values of $R = -0.414$ for unmitigated results, and $R = -0.388$ for mitigated results. These values, although lower than predicted, lie within general expectations since higher two-qubit gate error rates should correspond to lower levels of entanglement. Other factors that may impact negativity measurements include relaxation and dephasing time, crosstalk, and single-qubit gate errors.

We also plot the mean device negativity versus mean device CNOT error rate for all IBM Quantum devices in \cref{fig:devicecorrelation}, measuring $R = -0.591$ for unmitigated negativities and $R = -0.643$ for mitigated negativities. The results similarly lie within expectation, indicating the potential utility of our protocol as a scalable whole-device benchmarking tool.

We summarize all graph state experiment results in \cref{tab:summary}. In addition to tabulating device size, quantum volume, and mean device negativity, we display the sizes of the largest connected entanglement graphs with edges above a certain negativity threshold. In detail, columns with label $\mathcal{N}\geq x\%$ represent the size of the largest connected graph where edges exist only between qubits pairs whose measured negativity is at least $x\%$ of the maximum value. This metric allows us to simultaneously probe the scale and quality of clusters of entanglement. We observe a few general trends. Firstly, the standard deviation in negativities for each device typically decreases once we apply QREM. Similarly, the variance between mean device negativities also diminishes. This may be attributed to the bound on the maximum negativity and variance in mean readout error rates, which range from 1.1\% on \textit{ibm\texttt{\_}\!lagos} to 5.2\% on \textit{ibmq\texttt{\_}\!quito}. Secondly, for devices of similar size, quantum volume is not a good predictor of mean device negativity. For instance, \textit{ibm\texttt{\_}\!geneva}, which has a quantum volume of 32, has a mean device negativity of $0.461 \pm 0.089$ (QREM) compared to \textit{ibm\texttt{\_}\!kolkata}'s $0.407 \pm 0.134$, which has a quantum volume of 128. This may be ascribed to a couple of factors, the first being that quantum volume is defined over a subset of qubits, instead of the whole device, and the second being that our entanglement protocol utilizes primarily low-depth circuits. %We finally observe that the improvement in observed negativities due to QREM is significant and consistent. Without QREM, only a single device, \textit{ibm\texttt{\_}\!sherbrooke}, exhibits an $\mathcal{N}>90\%$ entanglement graph of any size at $\mathcal{N}=4$.

\begin{table*}
    \normalsize
    \centering
    \renewcommand{\arraystretch}{1.1}
    \begin{tabular}{c c c c c c c c}
        \hline\hline
        \multicolumn{8}{c}{\textbf{No QREM}}\\
        \hline
        Device & Qubits & QV & Mean $\mathcal{N}$ & $\mathcal{N}\geq 50\%$ & $\mathcal{N}\geq 75\%$ & $\mathcal{N}\geq 90\%$ & Whole-Device\\
        \hline
        \textit{lima} & 5 & 8 & $0.363 \pm 0.036$ & 5 & 3 & 0 & \greencheck\\
        \textit{belem} & 5 & 16 & $0.315 \pm 0.030$ & 5 & 0 & 0 & \greencheck\\
        \textit{quito} & 5 & 16 & $0.301 \pm 0.012$ & 5 & 0 & 0 & \greencheck\\
        \textit{manila} & 5 & 32 & $0.388 \pm 0.013$ & 5 & 5 & 0 & \greencheck\\ \hline
        \textit{jakarta} & 7 & 16 & $0.303 \pm 0.029$ & 7 & 0 & 0 & \greencheck\\
        \textit{oslo} & 7 & 32 & $0.403 \pm 0.019$ & 7 & 6 & 0 & \greencheck\\
        \textit{nairobi} & 7 & 32 & $0.375 \pm 0.005$ & 7 & 3 & 0 & \greencheck \\
        \textit{lagos} & 7 & 32 & $0.372 \pm 0.026$ & 7 & 3 & 0 & \greencheck\\
        \textit{perth} & 7 & 32 & $0.353 \pm 0.015$ & 7 & 0 & 0 & \greencheck\\ \hline
        \textit{guadalupe} & 16 & 32 & $0.335 \pm 0.027$ & 16 & 2 & 0 & \greencheck\\ \hline
        \textit{toronto} & 27 & 32 & $0.223 \pm 0.118$ & 9 & 3 & 0 & \greencheck\\
        \textit{geneva} & 27 & 32 & $0.239 \pm 0.118$ & 11 & 2 & 0 & \redx\ (26)\\
        \textit{hanoi} & 27 & 64 & $0.330 \pm 0.066$ & 26 & 3 & 0 & \greencheck\\
        \textit{auckland} & 27 & 64 & $0.372 \pm 0.065$ & 26 & 13 & 0 & \greencheck\\
        \textit{cairo} & 27 & 64 & $0.356 \pm 0.039$ & 27 & 4 & 0 & \greencheck\\
        \textit{mumbai} & 27 & 128 & $0.315 \pm 0.088$ & 23 & 4 & 0 & \greencheck\\
        \textit{montreal} & 27 & 128 & $0.247 \pm 0.061$ & 8 & 0 & 0 & \greencheck\\
        \textit{kolkata} & 27 & 128 & $0.333 \pm 0.118$ & 24 & 9 & 0 & \greencheck\\ \hline
        %\textit{brooklyn} & 65 && $\pm$ & & & &\\ \hline
        \textit{washington} & 127 & 64 & $0.290 \pm 0.117$ & 85 & 6 & 0 & \redx\ (121)\\ 
        \textit{sherbrooke} & 127 & 32 & $0.382 \pm 0.066$ & 125 & 76 & 4 & \greencheck\\ 
        \textit{brisbane} & 127 & - & $0.365 \pm 0.054$ & 125 & 26 & 0 & \greencheck\\ 
        \hline
        \textit{seattle} & 433 & - & $0.115 \pm 0.099$ & 11 & 3 & 0 & \redx\ (184)\\
        \hline
        \multicolumn{8}{c}{\textbf{QREM}}\\ \hline
        \textit{lima} & 5 & 8 & $0.470 \pm 0.011$ & 5 & 5 & 5 & \greencheck\\
        \textit{belem} & 5 & 16 & $0.427 \pm 0.010$ & 5 & 5 & 0 & \greencheck\\
        \textit{quito} & 5 & 16 & $0.486 \pm 0.010$ & 5 & 5 & 5 & \greencheck\\
        \textit{manila} & 5 & 32 & $0.487 \pm 0.003$ & 5 & 5 & 5 & \greencheck\\ \hline
        \textit{jakarta} & 7 & 16 & $0.482 \pm 0.007$ & 7 & 7 & 7 & \greencheck\\
        \textit{oslo} & 7 & 32 & $0.488 \pm 0.010$ & 7 & 7 & 7 & \greencheck\\
        \textit{nairobi} & 7 & 32 & $0.488 \pm 0.004$ & 7 & 7 & 7 & \greencheck\\
        \textit{lagos} & 7 & 32 & $0.466 \pm 0.008$ & 7 & 7 & 7 & \greencheck\\
        \textit{perth} & 7 & 32 & $0.482 \pm 0.011$ & 7 & 7 & 7 & \greencheck\\ \hline
        \textit{guadalupe} & 16 & 32 & $0.447 \pm 0.032$ & 16 & 16 & 11 & \greencheck\\ \hline
        \textit{toronto} & 27 & 32 & $0.403 \pm 0.075$ & 27 & 11 & 4 & \greencheck\\
        \textit{geneva} & 27 & 32 & $0.461 \pm 0.089$ & 26 & 26 & 25 & \greencheck\\
        \textit{hanoi} & 27 & 64 & $0.467 \pm 0.026$ & 27 & 27 & 17 & \greencheck\\
        \textit{auckland} & 27 & 64 & $0.437 \pm 0.060$ & 27 & 26 & 13 & \greencheck\\
        \textit{cairo} & 27 & 64 & $0.455 \pm 0.026$ & 27 & 27 & 9 & \greencheck\\
        \textit{mumbai} & 27 & 128 & $0.460 \pm 0.078$ & 27 & 27 & 23 & \greencheck\\
        \textit{montreal} & 27 & 128 & $0.424 \pm 0.055$ & 27 & 24 & 8 & \greencheck\\
        \textit{kolkata} & 27 & 128 & $0.407 \pm 0.134$ & 25 & 22 & 11 & \greencheck\\ \hline
        %\textit{brooklyn} & 65 && $\pm$ & & & & \\ \hline
        \textit{washington} & 127 & 64 & $0.408 \pm 0.102$ & 115 & 90 & 43 & \greencheck\\ 
        \textit{sherbrooke} & 127 & 32 & $0.472 \pm 0.023$ & 127 & 127 & 114 & \greencheck\\
        \textit{brisbane} & 127 & - & $0.467 \pm 0.048$ & 127 & 124 & 99 & \greencheck\\ 
        \hline
        \textit{seattle} & 433 & - & $0.340\pm0.118$ & 330 & 37 & 11 & \greencheck\ (active qubits)\\
        \hline\hline
    \end{tabular}
    \caption{Summary of bipartite negativities on IBM quantum devices. Negativities are acquired by performing parallelized quantum state tomography for every local qubit bipartition on the whole-device graph state. A maximally entangled pair has a negativity of 0.5. The table includes result with and without QREM, where the calibration matrices are obtained from tensoring single-qubit calibrations $A_i$. The column with label Mean $N$ shows the average device negativities. The columns with label $N \geq X\%$ represent the size of the largest connected graph with edges satisfying $N\geq X\%$ of the max negativity. A device is whole-device entangled if all qubits form a connected graph using edges with larger than zero negativity.\label{tab:summary}}
\end{table*}

\subsection{Generating 414-Qubit Graph States}
\begin{figure*}
    \centering
    \includegraphics[width=\textwidth]{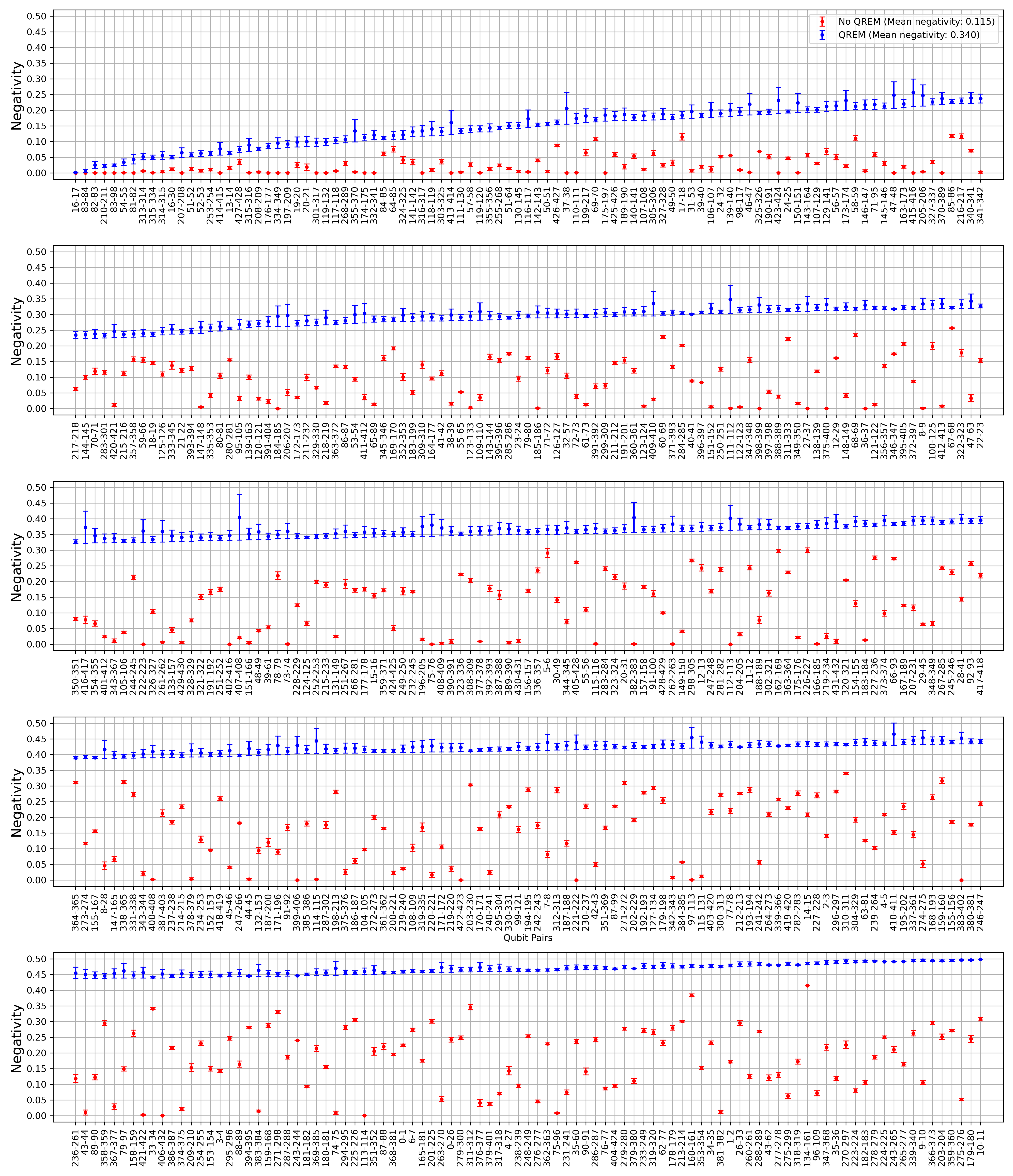}
    \caption{Negativity versus the corresponding qubit pairs where Bell states are projected to for the 433-qubit \textit{ibm\texttt{\_}\!seattle} device. Each negativity is averaged over all possible nearest-neighbour projections and trials, and the error bars represent $95\%$ confidence level calculated from two times the standard errors $(2\sigma)$ among the four trials. Red bars represent the results calculated from unmitigated counts and probability vectors, whereas blue bars represent the negativities calculated after applying QREM. The results are plotted in the ascending order of mitigated negativities. From this plot, we observe $\mathcal{N}>0$ for all nearest-neighbour qubit pairs.}
    \label{fig:433qBellStatenegativity}
\end{figure*}

\begin{figure*}
    \centering
    \includegraphics[width=0.95\textwidth]{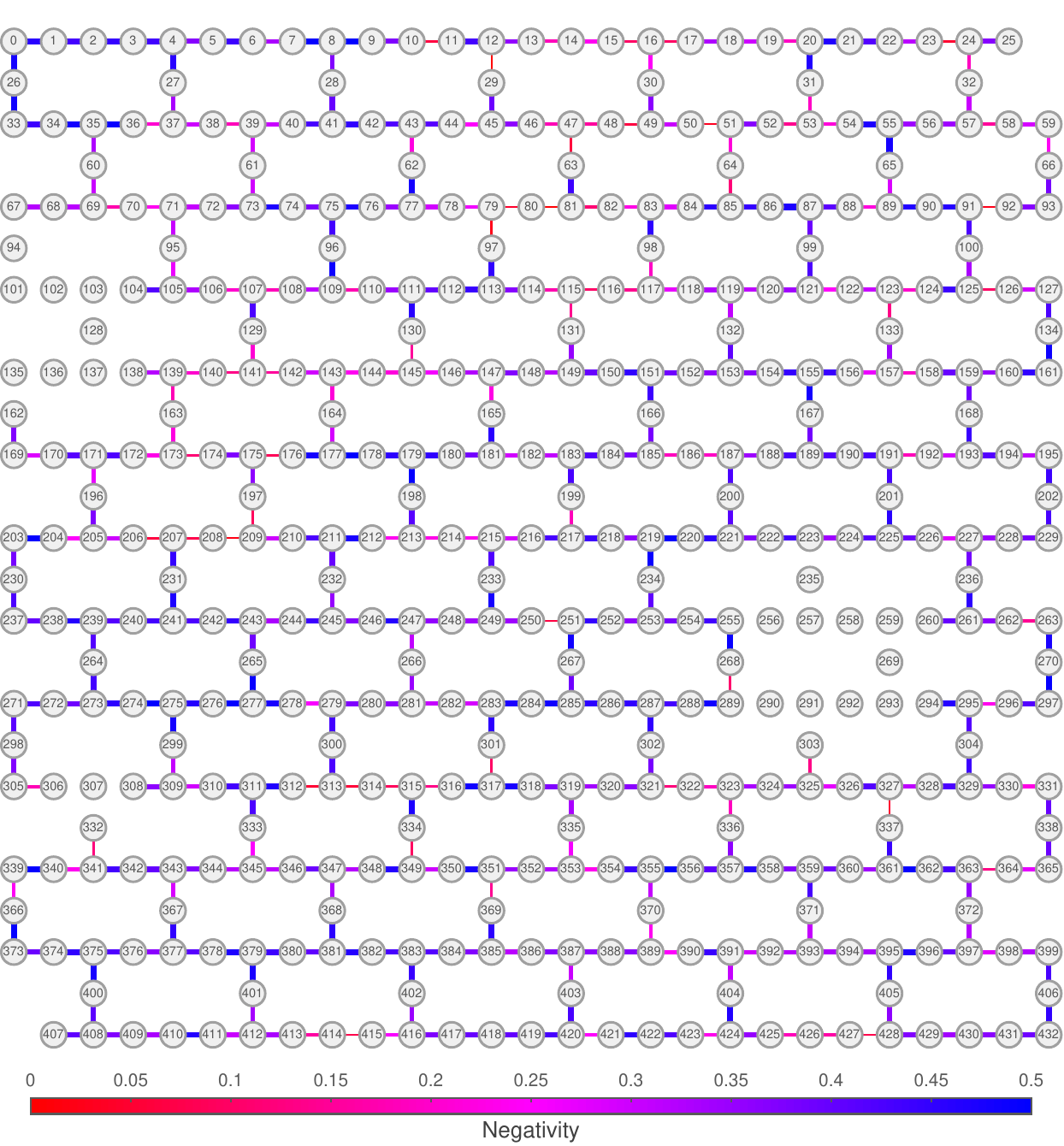}
    \caption{Graphic representation of entanglement with native-graph state prepared on \textit{ibm\_seattle} after QREM. Nodes represent the qubits, and edges represent their connections. The colour of an edge indicates the average negativity of the Bell states on the qubits incident to it. Some qubits are not connected (entangled) to the main graph since they are inactive and cannot be accessed.}
    \label{fig:ghznmap433qrem}
\end{figure*}

Using the same protocol, we characterize bipartite entanglement on a larger 414-qubit graph state prepared on the 433-qubit device $ibm\_seattle$, where at the time of experiment, 19 of the 433 device qubits were inoperable. \Cref{fig:433qBellStatenegativity} displays the average negativity versus nearest-neighbour qubit pairs. The average qubit pair negativity is found to be 0.115 without QREM, and 0.340 with QREM. We report proportionally higher readout error rates compared to previous devices. The coupling map is displayed in \cref{fig:ghznmap433qrem}. After mitigating for readout errors, all bipartitions not involving inactive qubits had measured negativities above 0.

\clearpage
\section{Preserving Whole-Device Graph States Via Dynamical Decoupling}\label{section:graphstatedecay}

We extend our study of preserving large-scale entanglement on IBM Quantum devices via dynamical decoupling to whole-device graph states. This application holds significant potential as graph states are considered to be more practically relevant than GHZ states on NISQ devices owing to their greater noise resilience. In particular, preserving large-scale entanglement in graph states is crucial for several measurement-based computation schemes, where qubits may experience long idle times due to the distributed manner in which quantum information is processed.

\begin{figure}[h]
    \centering
    \includegraphics[width=0.7\columnwidth]{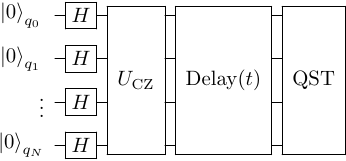}
    \caption{General circuit diagram for whole-device time-dependent negativity decay experiments. A variable circuit delay period is inserted between entangled state preparation and (parallel) quantum state tomography.\label{fig:graphstatedelay}}
\end{figure}

The methodology for testing dynamical decoupling on whole-device graph states parallels the GHZ case. A whole-device graph state is prepared as previously outlined. A variable delay period is inserted between state preparation and whole-device quantum state tomography. We compare negativity decays for circuits without mitigation (free decay), and circuits mitigated with PDD and Hahn echo. The general circuit diagram is shown in \cref{fig:graphstatedelay} and the PDD circuit in \cref{fig:circuitdelay}(b).

\subsection{Whole-Device Graphstate Negativity Decays on a 127-Qubit Device}
Time-dependent graphstate experiments are performed on the 127-qubit Eagle r3 backend \textit{ibm\texttt{\_}\!brisbane} with median $T_1 = 222.8$ \textmu s and $T_2 = 137.9$ \textmu s. The device did not have a published quantum volume at the time of experiment. We vary the delay period $t$ from 0 \textmu s up to 12 \textmu s in 1 \textmu s increments for free decays (idle qubits), Hahn echo, and PDD with 4 \textmu s\textsuperscript{-1} and 8 \textmu s\textsuperscript{-1} pulse rates (i.e. frequency of X-gates). We implement the double $\pi$-pulse with a 1:2:1 delay spacing. We perform single sets of experiments for each circuit delay value, executing all circuits with 4096 shots each. We apply readout error mitigation to all results.

\begin{figure*}
    \centering
    \includegraphics[width=\textwidth]{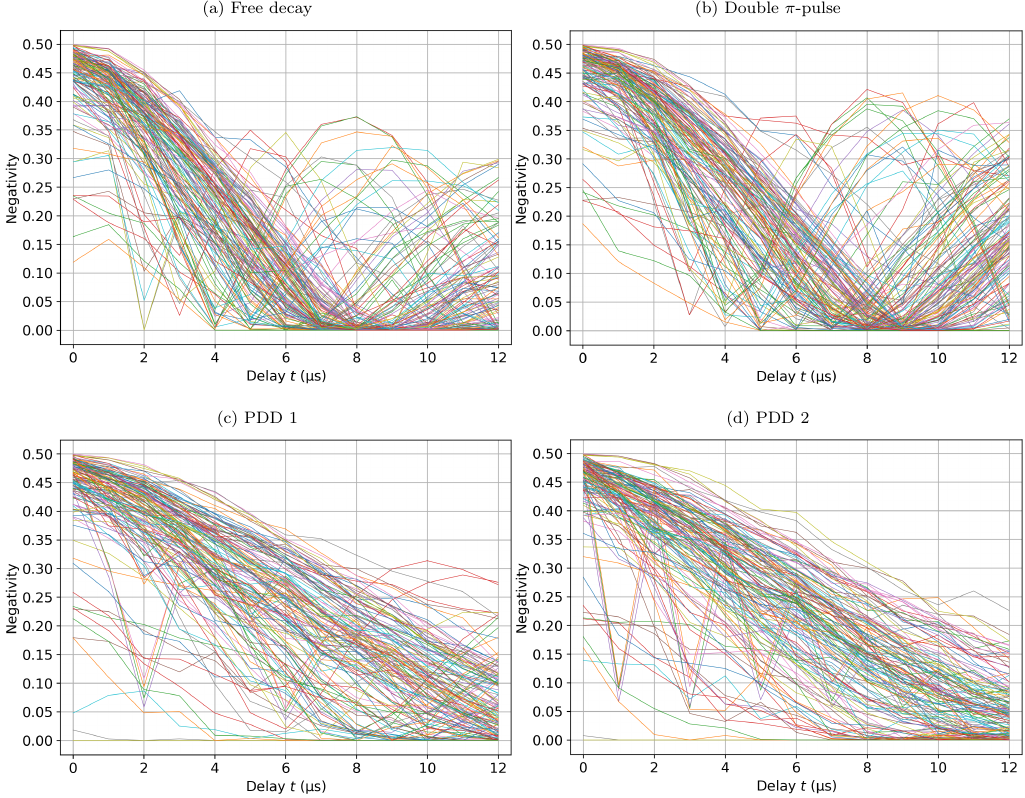}
    \caption{Negativity decays for all edges in a 127-qubit native graph state prepared on \textit{ibm\texttt{\_}\!brisbane} for \textbf{(a)} free decays, and \textbf{(b)} decays implementing double $\pi$-pulse, \textbf{(c)} periodic dynamical decoupling with 4 \textmu s$^{-1}$ pulse rate, and \textbf{(d)} periodic dynamical decoupling with 8 \textmu s$^{-1}$ pulse rate.\label{fig:negativitydecays}}
\end{figure*}

\begin{figure*}
    \centering
    \includegraphics[width=1\columnwidth]{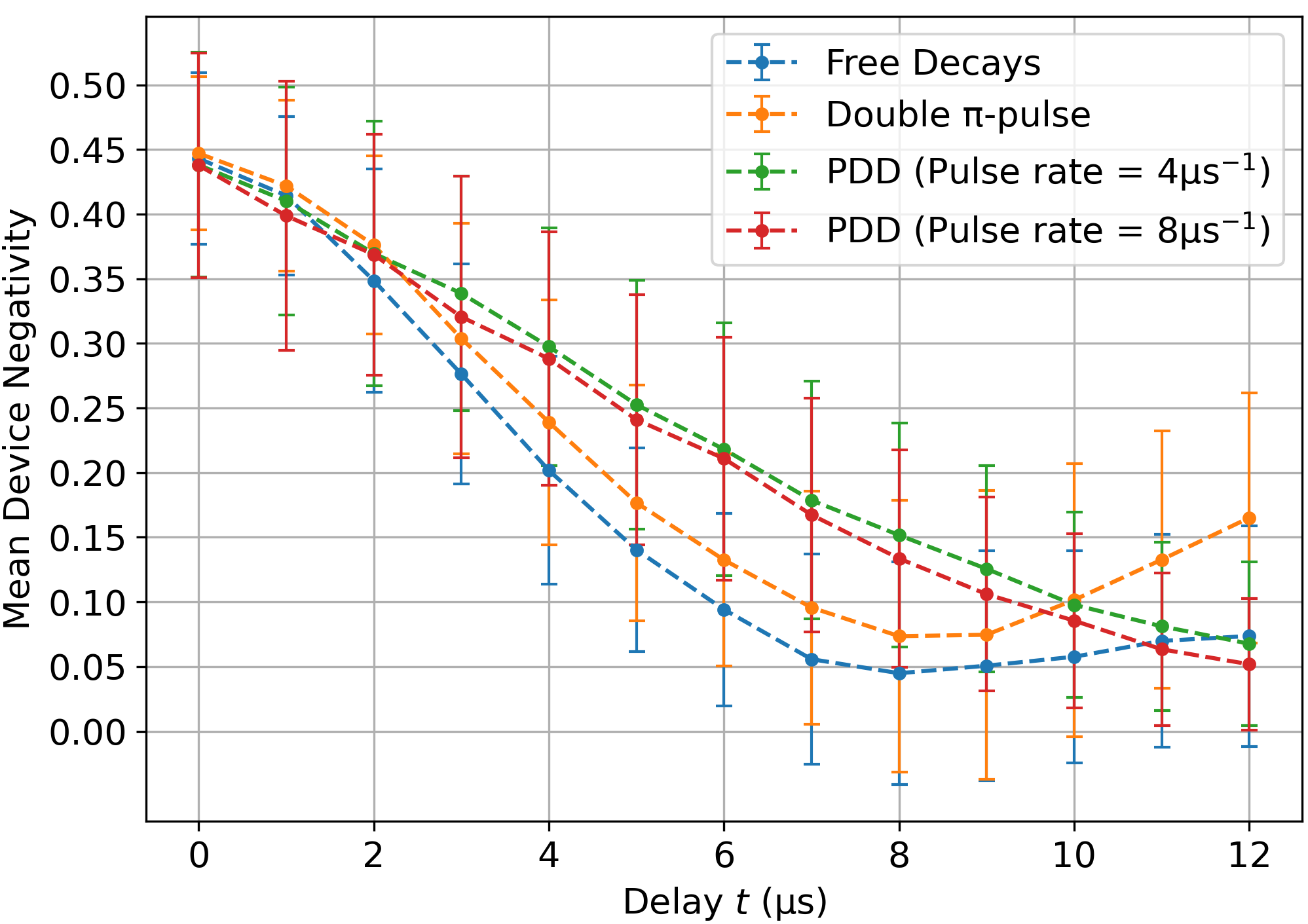}
    \caption{Mean device negativity decays corresponding to data shown in \cref{fig:negativitydecays}. Error bars represent the standard deviation between individual edge negativities.\label{fig:meannegativitydecays}}
\end{figure*}

\Cref{fig:negativitydecays} shows the negativity over time for each edge in the native-graph state, comparing free decays with various dynamical decoupling configurations. \Cref{fig:meannegativitydecays} displays the mean device negativity over time, where error bars represent the standard deviation between negativities of individual edges. 

Immediately apparent in \cref{fig:negativitydecays} are resurgent signals in negativity for several qubit pairs. Furthermore, in \cref{fig:meannegativitydecays}, we observe an average increase in device negativity of 0.025 for free decays, and 0.061 for double $\pi$-pulse, between $t=8$ \textmu s and $t=12$ \textmu s. We also note the sharp negativity oscillations of certain qubit pairs in PDD experiments. While we present data from only a single set of experiments, experiments performed shortly thereafter show similar oscillations in negativity for the same qubit pairs (with some device drift). These oscillations, while initially unexpected, are consistent with signals produced by residual ZZ interactions. 

ZZ interactions, also known as ZZ couplings or crosstalk are known to affect weakly anharmonic transmon qubits \cite{McKay2019Three-QubitBenchmarking,Magesan2020EffectiveGate,Ku2020SuppressionSystem,Sundaresan2020ReducingEchoes,Zhao2021SuppressionProcessor}. ZZ interactions represent a coherent noise process whose effect on idle qubits is effectively a controlled-phase rotation. Therefore, ZZ interactions can generate entanglement between qubit pairs, and conversely, accelerate disentanglement. Assuming no other noise mechanism such as dephasing is present, an isolated ZZ pair interaction produces a $|\cos(t)|$ signal in the negativity.

With this in mind, we consider the effects of dynamical decoupling sequences on the negativity decays of native-graph states. \Cref{fig:meannegativitydecays} shows that both PDD experiments demonstrate sizeable improvement in mean entanglement lifetimes over free decay and double $\pi$-pulse experiments. The double $\pi$-pulse experiment demonstrates slight improvement over free decays. In addition, increasing the PDD pulse rate does not appear to substantially improve the mean device negativity decay curve. From \cref{fig:negativitydecays}, we observe that PDD does not completely eliminate revivals in negativity, however does well to prolong the majority of pairwise negativity lifetimes and suppress some coherent noise artifacts.

While we focus on mean improvements in entanglement lifetimes for native graph states, recent results have shown that implementing a precisely timed dynamical decoupling sequence can more effectively cancel the coherent ZZ errors in a 12-qubit ring graph state \cite{Shirizly2023DissipativeQubits}. A similar approach of tailoring the dynamical decoupling sequence for native graph states may also be beneficial, although can be considerably more complex depending on the scale and connectivity of the underlying graph. For the purposes of improving the mean entanglement lifetime of a large-scale graph state, we show that even a simple PDD sequence incurs significant benefits. Additionally, we note that an adaption of our procedure may be potentially useful in detecting and characterizing the coherent noise effects.

\section{Discussion}

We prepared and studied several GHZ and native-graph states prepared across the range of IBM Quantum devices. In particular, we measured the time-dependent decay of entanglement in these states and verified the efficacy of dynamical decoupling in prolonging entanglement lifetimes.

For GHZ states, we developed a topology-agnostic circuit embedding algorithm that embeds $N$-qubit GHZ preparation circuits on heavy-hex quantum devices with least-depth $d \approx \sqrt{2N}$. Using the algorithm, we prepared a 32-qubit GHZ state on the 127-qubit \textit{ibm\texttt{\_}\!washington} device and measured a fidelity of $0.519 \pm 0.014$, after mitigating for readout errors via matrix-free measurement mitigation (M3). We demonstrated the efficacy of implementing dynamical decoupling-based techniques in preserving GHZ coherences on superconducting qubits. Specifically, we showed that incorporating either a \textpi -pulse or PDD substantially prolonged 7-qubit GHZ coherence times on the \textit{ibmq\texttt{_}\!mumbai} device. On \textit{ibm\texttt{_}\!hanoi}, we graphed the GHZ decoherence rate versus the state size $N$ up to $N=15$ qubits, fitting a linear trend of $\alpha=(7.13N+5.54)10^{-3}$ \textmu s$^{-1}$. This result supports the notion that IBM Quantum superconducting devices are naturally robust against superdecoherence.

For graph states, we developed a bipartite entanglement characterization protocol that constructs entanglement graphs depicting bipartite entanglement in IBM Quantum devices using as low as a constant 36 circuits. We used the protocol to verify and quantify whole-device bipartite entanglement over 20 different IBM Quantum systems, including three 127-qubit systems. We further showed entanglement across 414 qubits in a 433-qubit device. We then tested dynamical decoupling for preserving qubit pair negativities in a native-graph state prepared on the 127-qubit \textit{ibm\texttt{_}\!brisbane} device. We observed coherent noise signals consistent with residual ZZ interactions, which were partially suppressed after application of PDD. PDD led to an overall improvement in mean device bipartite entanglement lifetimes. We also note the potential utility of a running a similar procedure to detect and characterize the coherent noise signals.

Overall, our work highlights both some of the growing capabilities of NISQ devices alongside current limitations through the lens of large-scale entanglement. It also highlights the need for and benefit of noise mitigation and suppression techniques for generating and maintaining large-scale entanglement in NISQ devices.

\begin{acknowledgments}
This work was supported by the University of Melbourne through the establishment of an IBM Quantum Network Hub at the University. We would like to thank Seth Merkel for valuable discussions relating to transmon modelling.
\end{acknowledgments}

%\appendix
\bibliography{references}% Produces the bibliography via BibTeX.

%\clearpage
%\widetext
%\input{supplementary material}

\end{document}